\documentclass[aps,pra,reprint,twocolumn,groupedaddress]{revtex4-2}
\usepackage{graphicx}
\usepackage{dcolumn}
\usepackage{bm}
\usepackage{amsmath}
\usepackage{amssymb}
\usepackage{braket}
\usepackage{siunitx}
\usepackage[T1]{fontenc}
\usepackage{xcolor}
\usepackage{booktabs}

\usepackage{tikz}
\usetikzlibrary{decorations.pathmorphing,arrows.meta,calc,positioning}

\usepackage[colorlinks]{hyperref}
\usepackage[noabbrev]{cleveref}
\crefname{equation}{Eq.}{Eqs.}
\crefrangelabelformat{equation}{(#3#1#4--#5#2#6)}

\graphicspath{{.},{./fig/}}
\DeclareGraphicsExtensions{.pdf,.png,.eps}

\begin{document}

\preprint{APS/123-QED}

\title{Non-secular polariton leakage and dark-state protection in hybrid plasmonic cavities}

\author{Marco Vallone}
\email{marco.vallone@polito.it}
\affiliation{Dipartimento di Elettronica e Telecomunicazioni, Politecnico di Torino,\\
Corso Duca degli Abruzzi 24, 10129 Torino, Italy}

\date{\today}

\begin{abstract}
A major issue in exploiting plasmonic cavities as key components in nanotechnology is the effect of radiative and absorption losses on their electrodynamic behavior. Treating them as open-systems, we derive a time-local, completely positive master equation that retains non-secular interference between decay pathways and reduces to the standard secular description when the environment resolves polariton splitting. When it does not, the theory predicts order-one deviations from secular leakage dynamics, including bath-induced coherences and stabilization of dark polaritons, and provides a simple design criterion based on the ratio of polariton splitting to reservoir linewidth. A time-resolved leakage measurement, such as transmission, reflectivity, or photoluminescence, can be used to observe these effects.
\end{abstract}

\maketitle

\section{Introduction}
\label{s:introduction}

Surface plasmons in resonant cavities enable deep-subwavelength field confinement and enhanced light–matter interaction by reducing the effective mode volume. By tailoring the cavity geometry and the optical response of metallic or highly doped semiconductor reflectors, the coupling strength can be tuned from the weak to the strong and ultrastrong regimes, with corresponding modifications of radiative damping and plasmonic lifetimes upon hybridization~\cite{1988Raether,2012Biagioni_RPP,2015Palaferri_APL,2019Iqbal_PLASMONICS,2019Kockum_NATRP,2020Baranov_NATC,2021Yoo_NATP,2023WangZ_JAP}.

\begin{figure}[!t]
\centering
\includegraphics[width=1\columnwidth]{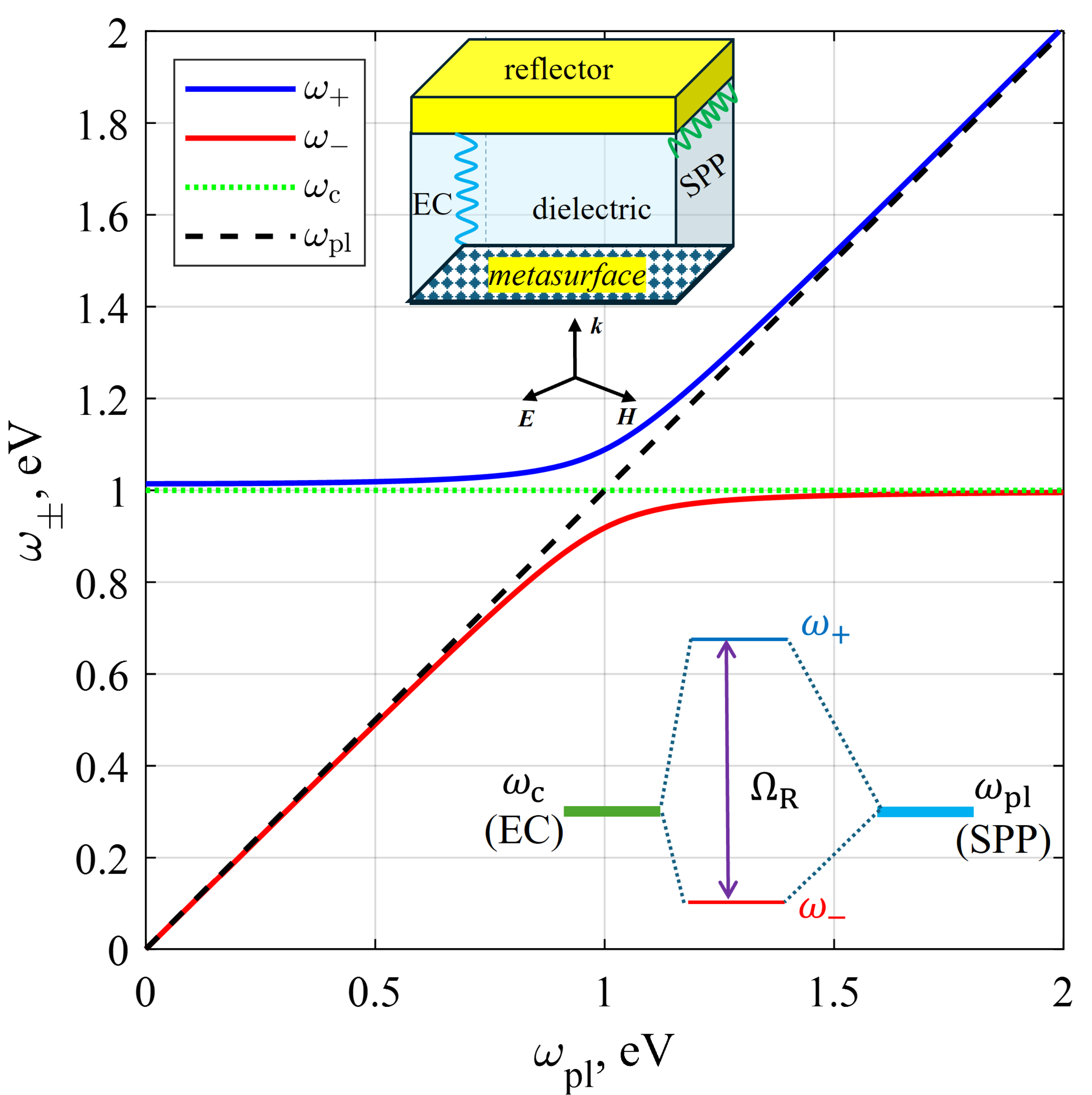}
\caption{Schematic representation of EC–SPP hybridization and dispersion of eigenfrequencies $\omega_\pm$. The cavity is illuminated from below and supports an EC mode at $\omega_\mathrm{c}$. SPPs with frequency $\omega_\mathrm{pl}$ are excited by a metasurface (e.g., a periodic lattice of nanoparticles) on the illuminated face of the cavity and propagate along the dielectric/reflector interface, which is assumed to be a medium-low loss material. The dispersion $\omega_\pm$ is calculated as in \cite{2025Vallone_NPJ} for $\omega_\mathrm{c} = 1$\,eV, plasma frequency $\Omega_\mathrm{pl} \approx 8.5$\,eV, dimensionless EC–SPP coupling constant $\xi = 0.02$, and zero losses.}
\label{f:fig_1}
\end{figure}

Periodic arrays of metallic nanoparticles or appropriate metasurfaces can efficiently excite surface plasmon polaritons (SPPs) at a reflector–dielectric interface of resonant cavities, as shown in Fig.\,\ref{f:fig_1}. These SPPs hybridize with electromagnetic cavity (EC) modes to form upper (UP) and lower (LP) polaritons with Hopfield-like dispersion~\cite{1958Hopfield_PR,2005Ciuti_PRB,2016Vasanelli_CRP,2020Baranov_NATC} and eigenfrequencies $\omega_\pm$. Such hybrid modes have been demonstrated from the terahertz to the visible range~\cite{1997Liu_PRB,2010Todorov_OE,2013Feuillet_SREP,2017Manceau_PRB,2019Vigneron_APL,2023Pisani_NATC}, providing an experimentally accessible platform to study driven and dissipative polariton dynamics.

A unified quantum open-system description for hybrid plasmonic cavities was developed in Ref.~\cite{2025Vallone_ARX} by combining a microscopic treatment of the dressed retarded photon propagator with a Gorini--Kossakowski--Sudarshan--Lindblad (GKSL) master equation~\cite{1976Lindblad_CMP,1976Gorini_JMP,2007Breuer} in the Markovian and secular (rotating-wave) limits. In that approach, the Dyson equation for the dressed propagator $D(\omega)$ in the random-phase approximation (RPA) yields a complex self-energy $\mathcal{S}(\omega)=\Sigma(\omega)-\mathrm{i}\Gamma(\omega)/2$ which, in the long-wavelength limit of the RPA, is proportional to the microscopic susceptibility $\chi(\omega)$ of the medium~\cite{1988Raether,2025Vallone_NPJ}. The real part $\Sigma(\omega)$ renormalizes the hybrid eigenfrequencies (including the observed blueshift of the UP and LP resonances in the strong-coupling regime~\cite{2020Baranov_NATC,2021Yoo_NATP,2023Pisani_NATC}), while the imaginary part sets absorption-induced linewidths $\Gamma(\omega)$ that enter the leakage terms; in the low-density and secular approximation this leads to linear rate equations for UP/LP populations and coherences with constant decay rates determined self-consistently from $\mathcal{S}(\omega)$.

In this work, we describe the system dynamics using a time-local, completely positive, and trace-preserving quantum master equation (QME) that retains non-secular interference (cross-damping) between decay pathways~\cite[Ch.~3]{2007Breuer}. The secular approximation keeps only dissipative contributions that couple transition operators with the same Bohr frequency and neglects cross terms oscillating as $\mathrm{e}^{\pm\mathrm{i}(\omega_i-\omega_j)t}$, which average out when $|\omega_i-\omega_j|$ is large compared to the relevant dissipative rates and the inverse bath correlation time (i.e., the spectral width of the environment).

Our formulation goes beyond the standard secular GKSL description and becomes essential in hybrid plasmonic cavities when the upper–lower polariton splitting $\Delta = \omega_+ - \omega_-$ is not resolved on the bath-linewidth scale $\gamma_\mathrm{D}$. In this regime, cross-damping makes the Kossakowski matrix nearly rank one, so the dissipation eigenchannels become collective bright and dark linear combinations of the underlying UP and LP transitions. Moreover, the dark combination can become quasi-protected~\cite{2008Chong_PRA,2023Kim_PRL,2025Bouteyre_OEX}, leading to large quantitative deviations from secular leakage models. The formalism we developed provides a simple, platform-relevant design criterion controlled by the ratio $\Delta/\gamma_\mathrm{D}$. Conversely, for $\Delta \gg \gamma_\mathrm{D}$, the cross terms are suppressed, the secular approximation is justified, and the QME generator reduces to the standard GKSL result.

It is important to note that the present theory applies to any resonant system in which a photonic mode interacts with bosonic matter excitations. This setting includes, for example, exciton–polaritons in semiconductor microcavities, intersubband polaritons in quantum wells, phonon–polaritons in polar dielectrics, two-level atoms interacting with an optical field, and plasmonic nanocavities as in the considered case \cite{2014LiuJ_OEX,2017LiuZ_SREP,2019Kockum_NATRP,2020Baranov_NATC,2021Yoo_NATP,2023Pisani_NATC,2023WangZ_JAP}.

The paper is organized as follows. In Sec.~\ref{s:model} we review the hybrid plasmonic cavity model and its EC--SPP Hopfield diagonalization, and we show how the Dyson equation yields complex polariton eigenfrequencies $\omega_\pm$ with shifts $\Sigma$ and linewidths $\Gamma$ set by the microscopic Drude--Lorentz susceptibility. In Sec.~\ref{s:non-secular} we formulate the driven--dissipative master equation in the polariton basis with a leakage dissipator constructed without the secular approximation and cast the dynamics in a vectorized Liouvillian form for numerical simulations. In Sec.~\ref{s:secular_breakdown} we compare the resulting non-secular dynamics with its secular approximation and discuss how bright and dark polaritons reveal interference effects and bath-induced coherence. We also suggest useful design criteria and approximate, analytical expressions for steady-state populations in the two compared descriptions, discussing their domain of validity. Section~\ref{s:conclusions} summarizes the results and outlines possible extensions. Throughout we set $\hbar=c=k_\mathrm{B}=1$; in the numerical results we express times in ps and frequencies in ps$^{-1}$.

\section{Hybrid plasmonic cavity and polaritonic modes}
\label{s:model}

Among the possible resonant cavities to which the formalism can be applied, we focus here on the hybrid plasmonic cavity schematically shown in Fig.~\ref{f:fig_1}, where an electromagnetic cavity mode at frequency $\omega_\mathrm{c}$ is strongly coupled to a surface plasmon polariton mode at frequency $\omega_\mathrm{pl}$, supported by a metallic or heavily doped semiconductor reflector.

In the Power–Zienau–Woolley multipolar picture~\cite{2012Todorov_PRB,2021Vukics_SREP}, the Hamiltonian describing the cavity as an isolated system without losses is given by
\begin{align}
\hat{H}_{\mathrm s} &=
\omega_{\mathrm c}\!\left(\hat a^{\dagger}\hat a+\tfrac{1}{2}\right)
+\omega_{\mathrm{pl}}\!\left(\hat{b}^{\dagger}\hat{b}+\tfrac{1}{2}\right) \nonumber \\
&-\frac{\mathrm{i}\xi\,\Omega_{\mathrm{pl}}}{2}
(\hat a-\hat a^{\dagger})(\hat{b}+\hat{b}^{\dagger})
+\frac{\xi^{2}\Omega_{\mathrm{pl}}^{2}}{2}
(\hat{b}+\hat{b}^{\dagger})^{2},
\label{eq:H_s_PZW}
\end{align}
where $\hat{a}$ and $\hat{b}$ are EC and SPP bosonic annihilation operators, respectively, $\Omega_{\mathrm{pl}}$ is the bulk plasma frequency, and $0\le\xi\le 1$ is a dimensionless coupling constant~\cite{2005Ciuti_PRB,2012Todorov_PRB,2016Vasanelli_CRP}. The quadratic term $(\hat{b}+\hat{b}^\dagger)^2$ is the multipolar analogue of the $\mathbf{A}^2$ term which arises from minimal coupling (where $\mathbf{A}$ is the electromagnetic vector potential) and is responsible for the experimentally observed blue shift of the polariton branches at strong coupling~\cite{2012Todorov_PRB,2019Kockum_NATRP,2020Baranov_NATC,2021Yoo_NATP,2023WangZ_JAP}.

Diagonalizing $\hat H_\mathrm{s}$ by a Hopfield--Bogoliubov transform~\cite{1958Hopfield_PR,2005Ciuti_PRB} yields bosonic polariton operators $\hat{B}_\pm$, which obey $[\hat{B}_k,\hat{B}_l^\dagger] = \delta_{kl}$, such that
\begin{equation}
\hat{H}_{\mathrm s} = \omega_{+} \hat{B}_{+}^\dagger \hat{B}_{+}
+ \omega_{-} \hat{B}_{-}^\dagger \hat{B}_{-},
\label{eq:H_s_polaritons}
\end{equation}
where $\omega_\pm$ are the UP and LP eigenfrequencies \cite{2025Vallone_NPJ}
\begin{equation}
    \omega_{\pm}^2 = 0.5(\omega_\mathrm{c}^2 + \widetilde{\omega}^2_\mathrm{pl}) \pm 0.5 \sqrt{\left(\omega_\mathrm{c}^2 - \widetilde{\omega}^2_\mathrm{pl}\right)^2 + 4 \Omega^2_\mathrm{pl} \omega_\mathrm{c}^2},
\label{eq:dispersion}
\end{equation}
and $\widetilde{\omega}_\mathrm{pl} = \sqrt{\omega^2_\mathrm{pl} + \Omega^2_\mathrm{pl}}$.

In our previous work \cite{2025Vallone_ARX} we showed that, at the resonant crossing $\omega_\mathrm{pl} = \omega_\mathrm{c}$, the effect of the electronic medium and material losses on the cavity photon field can be equivalently described by the retarded photon propagator $D(\omega)$, which satisfies a Dyson equation of the form $D^{-1}(\omega) = D_0^{-1}(\omega) - \mathcal{S}(\omega)$, where $D_0(\omega)$ is the free cavity photon propagator. This formalism leads to an expression for $\omega_\pm$ given by
\begin{equation}
\omega_\pm = \omega_\mathrm{c} \pm \zeta + \Sigma - \mathrm{i}\Gamma/2,
\label{eq:eigenfreq}
\end{equation}
where $2\zeta$ quantifies the mode frequency separation (the Rabi frequency), $\Sigma = \Re[\mathcal{S}]$ is the energy renormalization, and $\Gamma = -2\Im[\mathcal{S}]$ defines the linewidth and depends on the losses due to absorption in the cavity (and possibly on radiative losses). In the long-wavelength limit of the RPA, the self-energy $\mathcal{S}$ is proportional to the complex electric susceptibility $\chi$ of the medium,
\begin{equation}
\chi(\omega) = \frac{\Omega_{\mathrm{pl}}^2}
{\omega_{\mathrm{pl}}^2 - \omega^2 - \mathrm{i}\gamma_\mathrm{D}\omega},
\label{eq:chi_Drude}
\end{equation}
where the rate $\gamma_\mathrm{D}$ accounts for SPP absorption in the medium \cite{1988Raether,2012Biagioni_RPP,2025Vallone_NPJ}. Except for $\omega_\mathrm{c}$, all quantities in \cref{eq:eigenfreq} depend explicitly on the losses, and their functional forms are briefly recalled in Appendix\,\ref{app:eigenfrequencies}.

\section{Driven--dissipative dynamics}
\label{s:non-secular}

The hybrid cavity can be conveniently described as a bosonic two-mode system, where the Fock states $\ket{n_{+}, n_{-}}$, with $ n_\pm = 0, 1, 2, \ldots$, form an orthonormal basis for the Hilbert space of the isolated system (the vacuum corresponds to $n_{+} = n_{-} = 0$). These states are generated from the vacuum by the upper (UP, $+$) and lower (LP, $-$) polariton creation operators $\hat{B}_\pm^\dagger$, 
%
%
%
and the corresponding number operators are $\hat{N}_\pm = \hat{B}_\pm^\dagger \hat{B}_\pm$. The Hamiltonian $\hat{H}_\mathrm{s}$ of the isolated system, \cref{eq:H_s_polaritons}, is diagonal in this basis, with eigenvalues $E_\mathbf{n} \equiv E_{n_{+}, n_{-}} = n_{+} \omega_{+} + n_{-} \omega_{-}$.

To investigate the cavity dynamics, we consider a time-dependent Hamiltonian of the form
\begin{equation}
    \hat{H}(t) = \hat{H}_\mathrm{s} + \hat{H}_\mathrm{drive}(t),
    \label{eq:H_total}
\end{equation}
where $\hat{H}_\mathrm{drive}(t)$ accounts for external coherent drives. We model two types of coherent excitation: a cavity-like drive at frequency $\omega_\mathrm{c}$ that populates the polariton modes, and a Raman-like drive with frequency $\omega_\mathrm{ext}$ close to the UP--LP separation $\Delta$, which coherently couples the two branches. They can be written in the Schr\"odinger picture as
\begin{align}
    \hat{H}_\mathrm{drive}(t)
    &= \sum_{k=\pm} \Bigl[ f_k \, e^{-\mathrm{i}\omega_\mathrm{c} t} \hat{B}_k^\dagger
    + f_k^* \, e^{\mathrm{i}\omega_\mathrm{c} t} \hat{B}_k \Bigr] \nonumber \\
    &\quad + \mathcal{R}_\mathrm{UP\!-\!LP} \cos(\omega_\mathrm{ext} t)
    \bigl( \hat{B}_+^\dagger \hat{B}_- + \hat{B}_-^\dagger \hat{B}_+ \bigr),
    \label{eq:H_drive}
\end{align}
where $f_\pm$ are complex drive amplitudes determined by the overlap between the cavity field and the polariton modes, and $\mathcal{R}_\mathrm{UP\!-\!LP}$ controls the strength of the Raman-like interbranch coupling. Both drives can be switched on and off according to the designed simulation procedure. We also define the detuning $\delta = \Delta - \omega_\mathrm{ext}$, although we usually focus on the resonant case $\delta = 0$ when discussing dark and bright modes in Sec.~\ref{s:secular_breakdown}.

We derive the dissipative generator for the undriven Hamiltonian $\hat{H}_\mathrm{s}$ and include the external drive $\hat{H}_\mathrm{drive}(t)$ \textit{a posteriori} in the QME. The resulting evolution is written in the laboratory frame, and the interaction picture with respect to $\hat{H}_\mathrm{s}$ is used only in the intermediate steps of the Born–Markov construction. This choice ensures that the leakage rates depend only on the Bohr spectrum of $\hat{H}_\mathrm{s}$ and the bath spectral density.

To account for dissipation, we consider four physical mechanisms: (i) leakage (radiative and/or absorptive) of the polariton modes, (ii) incoherent scattering from the upper to the lower polariton (UP$\to$LP), (iii) incoherent scattering from the lower to the upper polariton (LP$\to$UP), and (iv) pure dephasing of each polariton branch.

\subsection{Leakage: mode absorption and radiative losses}

The undriven Hamiltonian for the open system is written as $\hat{H}_{\mathrm{s}} + \hat{H}_{\mathrm{env}} + \hat{H}_{\mathrm{int}}$, where $\hat{H}_{\mathrm{env}}$ is the environment (or bath) Hamiltonian and $\hat{H}_{\mathrm{int}}$ describes the system-bath interactions. 
We assume that the dominant leakage mechanism for both polariton branches is the interaction with a common electromagnetic reservoir, such as radiative leakage through the cavity boundaries or absorption in the reflector layer, therefore $\hat{H}_{\mathrm{int}}$ can be written as
\begin{equation}
    \hat{H}_{\mathrm{int}}
    = \hat{X} \otimes \hat{\mathcal{B}},
    \qquad
    \hat{X} = \hat{B}_+ + \hat{B}_- + \hat{B}_+^\dagger + \hat{B}_-^\dagger,
    \label{eq:H_int_BR}
\end{equation}
where $\hat{\mathcal{B}}$ is a bath operator that annihilates collective excitations in the electronic medium and in the external radiation continuum.

Following the standard Bloch–Redfield (BR) formalism~\cite{2007Breuer,2020Hartmann_PRA,2016Eastham_PRA,2024Pradilla_PRA}, we expand $\hat{X}$ in terms of eigenoperators of $\hat{H}_\mathrm{s}$,
\begin{equation}
    \hat{X} = \sum_j A_j \hat{\sigma}_j + A_j^* \hat{\sigma}_j^\dagger,
    \qquad
    \hat{\sigma}_j = \ket{n_j}\bra{m_j},
\end{equation}
where $\ket{n_j}$ and $\ket{m_j}$ are eigenstates of $\hat{H}_\mathrm{s}$, and each $\hat{\sigma}_j$ is associated with a well-defined Bohr frequency $\omega_j = E_{m_j} - E_{n_j}$. Assuming the Fock states include no more than $N_\mathrm{max}$ excitations, they span a truncated Hilbert space $\mathcal{H}$ of dimension $d = (N_\mathrm{max} + 1)^2$, and the relevant transitions connect the vacuum to single-polariton states or, in higher sectors, differ by one excitation in either branch.

The bath is characterized by the correlation function
$C(t)=\langle \hat{\mathcal B}(t)\hat{\mathcal B}(0)\rangle$
and by a rate-valued spectral density $J(\omega)$, related by the
fluctuation--dissipation theorem to the dissipative part of the susceptibility
\cite{1951Callen_PR,1966Kubo_RPP,2012Weiss}. Following Ref.~\cite{2024Pradilla_PRA},
we model $J(\omega)$ as a Lorentzian centered at the SPP resonance
$\omega_{\rm pl}$ with full width at half maximum (FWHM) $\gamma_D$,
\begin{equation}
J(\omega)=\Gamma_{\rm pl}\,
\frac{(\gamma_D/2)^2}{(\omega-\omega_{\rm pl})^2+(\gamma_D/2)^2},
\label{eq:J_Lorentz}
\end{equation}
where $\Gamma_{\rm pl}\equiv J(\omega_{\rm pl})$ sets the peak leakage scale.
The width $\gamma_D$ controls the bath correlation time (and hence the suppression
of cross-damping terms), while $\Gamma_{\rm pl}$ fixes the overall magnitude of the
secular decay rates $\Gamma_\pm\equiv J(\omega_\pm)$; in this effective description
they are independent parameters. On exact resonance, $\omega_\mathrm{pl}=\omega_\mathrm{c}$ and
$\omega_\pm=\omega_\mathrm{pl}\pm\Delta/2$, so the Lorentzian symmetry implies
\begin{equation}
    \Gamma_\pm \equiv \Gamma = \Gamma_{\rm pl}/\!\bigl[1+(\Delta/\gamma_D)^2\bigr]. 
\label{eq:Gamma_pm_def}
\end{equation}
In the numerical simulations below, we set $\Gamma_{\rm pl}=2\gamma_D$ to fix the overall leakage timescale; this choice rescales absolute decay rates without reducing the generality. 

Within the Born–Markov approximation, the BR master equation for the reduced density operator $\hat{\rho}(t)$ takes the form
\begin{equation}
    \frac{d\hat{\rho}}{dt}
    = -\mathrm{i}\bigl[\hat{H}_\mathrm{s},\hat{\rho}\bigr]
    + \mathcal{K}_\mathrm{BR}\hat{\rho},
    \label{eq:BR_general}
\end{equation}
where the dissipative kernel $\mathcal{K}_\mathrm{BR}$ can be expressed in the basis of transition operators $\{\hat{\sigma}_j\}$ as \cite{2007Breuer,2024Pradilla_PRA}
\begin{align}
    \mathcal{K}_\mathrm{BR}\hat{\rho}
    &= \sum_{i,j}
    \Bigl\{
    -\mathrm{i}
    \Bigl[
        \Lambda_{ij}(\omega_j)\,\hat{\sigma}_i^\dagger\hat{\sigma}_j \hat{\rho}
        - \hat{\rho}\,\Lambda_{ij}(\omega_i)\,\hat{\sigma}_i^\dagger\hat{\sigma}_j
    \Bigr]
    \nonumber \\
    &\quad + \mathrm{i}
    \bigl[\Lambda_{ij}(\omega_j)-\Lambda_{ij}(\omega_i)\bigr]\,
    \hat{\sigma}_j \hat{\rho}\hat{\sigma}_i^\dagger \nonumber \\
    &\quad - \tfrac{1}{2}
    \Bigl[W_{ij}(\omega_j)\,\hat{\sigma}_i^\dagger\hat{\sigma}_j \hat{\rho}
    + \hat{\rho}\,W_{ij}(\omega_i)\,\hat{\sigma}_i^\dagger\hat{\sigma}_j\Bigr]
    \nonumber \\
    &\quad + \tfrac{1}{2}
    \bigl[W_{ij}(\omega_j)+W_{ij}(\omega_i)\bigr]\,
    \hat{\sigma}_j \hat{\rho}\hat{\sigma}_i^\dagger
    \Bigr\}.
    \label{eq:BR_kernel}
\end{align}
Here, $W_{ij}(\omega)$ and $\Lambda_{ij}(\omega)$ are the decay and Lamb-shift matrices, respectively, given by $W_{ij}(\omega) = A_i^* A_j\,\gamma(\omega)$ and $\Lambda_{ij}(\omega) = A_i^* A_j\,\lambda(\omega)$, where the scalar functions $\gamma(\omega)$ and $\lambda(\omega)$ are determined by $J(\omega)$ as described in \cite{2007Breuer}, 
\begin{align}
\label{eq:gamma}
    \gamma(\omega)
    &= 2\pi\,\theta(\omega)\,J(\omega),\\
\label{eq:lambda}
    \lambda(\omega)
    &= \mathcal{P}\!\int_0^{\infty} d\omega'\,J(\omega')
    \left(
        \frac{1}{\omega-\omega'} + \frac{1}{\omega+\omega'}
    \right), 
\end{align}
in the zero temperature limit. This is a valid assumption, since the bath is approximately in the vacuum state or in a thermal state such that $k_B \ll \mathrm{min}_i(\omega_i)$, as commonly assumed in quantum optics \footnote{More specifically, we define $\int_0^\infty \mathrm{d}\tau\, \mathrm{e}^{\mathrm{i}\omega \tau} C(\tau) = \tfrac12 \gamma(\omega) + \mathrm{i} \lambda(\omega)$. For a bosonic bath with spectral density $J(\omega)$ in thermal equilibrium, $C(t) = \int_0^\infty \mathrm{d}\omega' J(\omega') \{ [n(\omega')+1]\mathrm{e}^{-\mathrm{i}\omega' t} + n(\omega')\mathrm{e}^{\mathrm{i}\omega' t} \}$, where $n(\omega)$ is the Bose distribution. This yields $\gamma(\omega) \propto J(\omega) [n(\omega)+1]$ for $\omega>0$ and $\gamma(\omega) \propto J(|\omega|) n(|\omega|)$ for $\omega<0$, while $\lambda(\omega)$ is a principal-value integral over $J(\omega)$ (its Hilbert transform). At zero temperature, \cref{eq:gamma,eq:lambda} follow.}. 
It should also be noted, for clarity, that the bath operator $\hat{\mathcal{B}}$ appears only through the scalar functions $\gamma(\omega)$ and $\lambda(\omega)$ once the bath degrees of freedom have been traced out.\footnote{Inserting $\hat{H}_{\mathrm{int}}(t) = \hat{X}(t) \otimes \hat{\mathcal{B}}(t)$ into the Born–Markov master equation and tracing over the environment, characterized by a density operator $\hat{\rho}_\mathrm{env}$, yields bath correlation functions $C(\tau) = \mathrm{Tr}_\mathrm{env}[\hat{\mathcal{B}}(\tau)\hat{\mathcal{B}}(0)\hat{\rho}_\mathrm{env}]$ multiplied by products of system operators $\hat{X}(t)$ and $\hat{X}(t-\tau)$~\cite{2007Breuer}. Thus, the reservoir operator $\hat{\mathcal{B}}$ disappears from the reduced dynamics, surviving only through the scalar functions $\gamma(\omega)$ and $\lambda(\omega)$ obtained from $C(\tau)$ (or equivalently from the spectral density $J(\omega)$ associated with the Drude–Lorentz susceptibility), while the Bloch–Redfield tensors $\Gamma_{ij}(\omega)$ and $\Lambda_{ij}(\omega)$ arise from combining these scalar kernels with the system coefficients $A_i^* A_j$ that weight the transition operators $\hat{\sigma}_j$.}

The key point of this approach is that both $W_{ij}(\omega)$ and $\Lambda_{ij}(\omega)$ inherit the full frequency dependence of $J(\omega)$, and therefore depend explicitly on the UP and LP frequencies $\omega_\pm$ and their separation $\Delta$, describing the dynamics without invoking the secular approximation and including all possible couplings between populations and coherences.

However, the generator $-\mathrm{i}\bigl[\hat{H}_\mathrm{s},\cdot\bigr]
+ \mathcal{K}_\mathrm{BR}$ is not in Lindblad form and does not guarantee complete positivity of $\hat{\rho}(t)$. To obtain a Lindblad-like, completely positive master equation that preserves the non-secular physics captured by the BR equation, we follow the prescription proposed in \cite{2024Pradilla_PRA}, symmetrizing the frequency-dependent coefficients using an arithmetic mean for the Lamb-shift terms and a geometric mean for the dissipative terms:
\begin{subequations}
\label{eq:aL_gG_means}
\begin{align}
    \widetilde{\Lambda}_{ij}
    &= \frac{\Lambda_{ij}(\omega_i) + \Lambda_{ij}(\omega_j)}{2},
    \label{eq:aL} \\
    \widetilde{W}_{ij}
    &= \sqrt{W_{ij}(\omega_i)\,W_{ij}(\omega_j)} = \sqrt{J(\omega_i)J(\omega_j)}.
    \label{eq:gG}
\end{align}
\end{subequations}
Hereafter we absorb all frequency-independent prefactors (including the relevant system matrix elements) into the effective rate function $J(\omega)$, so that $W_{ii}(\omega_i)\equiv J(\omega_i)$ and the geometric-mean prescription can be written as in Eq.~\eqref{eq:gG}.

The first choice ensures that the energy correction can be reorganized into a commutator with a Hermitian Lamb-shift Hamiltonian, while the second preserves the frequency dependence of the decay rates and strongly suppresses off-diagonal couplings when the corresponding Bohr frequencies are far apart compared to the bath linewidth.

Using the replacements~\cref{eq:aL_gG_means} in \cref{eq:BR_kernel}, the QME can be written as
\begin{equation}
    \frac{d\hat{\rho}}{dt}
    = -\mathrm{i}\bigl[\hat{H}_\mathrm{s} + \hat{H}_\mathrm{LS},\hat{\rho}\bigr]
    + \mathcal{L}_{a\Lambda\text{--}gW}\hat{\rho},
    \label{eq:AG_master}
\end{equation}
with the Lamb-shift Hamiltonian
\begin{equation}
    \hat{H}_\mathrm{LS}
    = \sum_{i,j} \widetilde{\Lambda}_{ij}\,\hat{\sigma}_i^\dagger\hat{\sigma}_j
    \label{eq:AG_LS}
\end{equation}
and the dissipator
\begin{equation}
    \mathcal{L}_{a\Lambda\text{--}gW}\hat{\rho}
    = \sum_{i,j} \widetilde{W}_{ij}
    \left(
        \hat{\sigma}_j \hat{\rho}\hat{\sigma}_i^\dagger
        - \tfrac{1}{2}\bigl\{\hat{\sigma}_i^\dagger\hat{\sigma}_j,\hat{\rho}\bigr\}
    \right) .
    \label{eq:AG_Diss}
\end{equation}
\cref{eq:AG_master,eq:AG_LS,eq:AG_Diss} are in GKSL form provided that the dissipative kernel $\widetilde{W}_{ij}$ is positive semidefinite. Since the BR construction may produce a Hermitian Kossakowski matrix $\widetilde{W}$ with small negative eigenvalues, we diagonalize $\widetilde{W}=U D U^\dagger$, with $U$ unitary and $D=\mathrm{diag}(d_1,\dots,d_M)$ real, and enforce complete positivity by projecting onto the positive part: $d_\mu^{(+)}=\max(d_\mu,0)$ and $\widetilde{W}_+=U D^{(+)} U^\dagger$, where $D^{(+)}=\mathrm{diag}(d_1^{(+)}, \dots, d_M^{(+)})$. In the truncated polariton space used in Section\ref{s:secular_breakdown}, the relevant leakage kernel reduces to a $2\times2$ matrix, whose eigenvalues (rates) and eigenvectors (jump operators) are given explicitly in that Section (and in Appendix~\ref{app:component_odes}).

Introducing the jump operators $\hat Q_\mu=\sum_j U_{j\mu}\hat\sigma_j$, the master equation becomes
\begin{subequations}
\label{eq:AG_plus_L_leak}
\begin{align}
    \frac{d\hat{\rho}}{dt}
    &= -\mathrm{i}\bigl[\hat{H}_\mathrm{s} + \hat{H}_\mathrm{LS},\hat{\rho}\bigr]
    + \mathcal{L}_\mathrm{leak}\,\hat{\rho} 
    \label{eq:AG_plus}  \\
\mathcal{L}_\mathrm{leak} &= \sum_\mu d_\mu^{(+)} \,\mathcal{D}[\hat{Q}_\mu]
    \label{eq:L_leak}
\end{align}
\end{subequations}
where $\mathcal{D}[\hat{O}]\,\hat{\rho}=\hat{O}\hat{\rho}\hat{O}^\dagger-\tfrac12\{\hat{O}^\dagger\hat{O},\hat{\rho}\}$ for any operator $\hat{O}$. 
Diagonalizing the Hermitian Kossakowski matrix $\widetilde{W}$ only amounts to a rotation in the space of leakage operators, and the possible cross-damping encoded in the off-diagonal terms of $\widetilde{W}$ is not lost, but is absorbed into the jump operators $\hat Q_\mu$. This leads to the dissipator $\mathcal{L}_\mathrm{leak}$, whose explicit form is $\sum_{j\mu} d_\mu^{(+)} \,\mathcal{D}[U_{j\mu}\hat\sigma_j]$ (with $d_\mu\to d_\mu^{(+)}$ because of the positivity projection, if needed).
A detailed assessment of this procedure and its accuracy is given in Ref.~\cite{2024Pradilla_PRA}.

\subsection{UP-LP scattering and pure dephasing}
\label{s:UP-LP_scattering}

Incoherent UP$\leftrightarrow$LP scattering processes are described by the jump operators
\begin{equation}
    \hat{L}_\downarrow = \hat{B}_-^\dagger \hat{B}_+,
    \qquad
    \hat{L}_\uparrow   = \hat{B}_+^\dagger \hat{B}_-,
    \label{eq:Lscat}
\end{equation}
with corresponding rates $\gamma_\downarrow$ and $\gamma_\uparrow$, respectively, which may be constrained by detailed-balance relation if the baths are thermal. Pure dephasing of each polariton branch is modeled by the number operators
\begin{equation}
    \hat{L}_{\phi,+} = \hat{N}_+,
    \qquad
    \hat{L}_{\phi,-} = \hat{N}_-,
    \label{eq:Ldeph}
\end{equation}
with dephasing rates $\gamma_{\phi,+}$ and $\gamma_{\phi,-}$, which we assume to be equal and denote as $\gamma_{\phi}$.

\subsection{Final form of the Lindbladian master equation}
Collecting all contributions, the full driven dynamics of the hybrid cavity, including the coherent drives according to \cref{eq:H_drive} and additional scattering and dephasing channels, is described by the non-secular master equation
\begin{equation}
    \frac{d\hat{\rho}}{dt}= -\mathrm{i}\bigl[\hat{H}_\mathrm{s} + \hat{H}_\mathrm{LS} + \hat{H}_\mathrm{drive}(t),\hat{\rho}\bigr]  + \mathcal{L}_\mathrm{tot}\,\hat{\rho},
    \label{eq:ME_total_AG}
\end{equation}
where
\begin{align}
    \mathcal{L}_\mathrm{tot} &= \mathcal{L}_\mathrm{leak} + \mathcal{L}_\mathrm{scatt} + \mathcal{L}_\mathrm{\phi} \nonumber \\
    &= \mathcal{L}_\mathrm{leak}+  \sum_{k=\downarrow,\uparrow}\gamma_k\,\mathcal D[\hat{L}_k] +  \gamma_{\phi} \sum_{k=\pm}\mathcal D[\hat{N}_k]
    \label{eq:Ltotal}
\end{align}
is the total dissipator. Here $\mathcal{L}_\mathrm{leak}$, \cref{eq:L_leak}, encodes the leakage due to the interactions with the bath (shortly, the bath leakage, i.e., mode absorption and radiative losses), while $\mathcal{L}_\mathrm{scatt} = \sum_{k=\downarrow,\uparrow}\gamma_k\,\mathcal D[\hat{L}_k]$ and $\mathcal{L}_\phi = \gamma_{\phi} \sum_{k=\pm}\mathcal D[\hat{N}_k]$ describe UP$\leftrightarrow$LP scattering and pure dephasing, respectively.

Equations~\eqref{eq:ME_total_AG} and \eqref{eq:Ltotal} provide an accurate Lindblad (GKSL) description of the leakage dynamics of the hybrid cavity: the generator is linear in $\hat{\rho}$ and defines a completely positive and trace-preserving dynamical map by construction of the Kossakowski matrix, retaining the non-secular character of the leakage dynamics through the Hermitian quadrature coupling $\hat X$. In particular, the off-diagonal elements of $\widetilde{W}_+$ remain nonzero whenever the corresponding Bohr frequencies are closer than the bath linewidth, so that different polariton transitions still interfere in the dissipator. This goes beyond the standard secular Lindblad treatment, where the Kossakowski matrix is diagonal in the Bohr-frequency basis and all such cross terms are discarded.

We emphasize that, among all dissipative channels, only the leakage term is modified by retaining non-secular contributions: the Hermitian jump operator $\hat X$ generates cross-terms with oscillating factors at frequencies $\pm(\omega_{+}\pm\omega_{-})$, which can be discarded in the secular approximation. By contrast, the other considered dissipative channels does not contain similar cross-terms.


\cref{eq:ME_total_AG} can be written explicitly as a system of first-order linear differential equations for a vectorized density matrix. To this end, we stack the columns of $\hat{\rho}(t)$ into a single $d^2$-component vector, $\mathrm{vec}(\hat{\rho}(t))$, via the map $\mathrm{vec} : \mathbb{C}^{d \times d} \to \mathbb{C}^{d^2}$. Using the identity~\cite{Simpson2017,2025GU_PRE,2015AmShallem_ARX}
\begin{equation}
    \mathrm{vec}(\hat A\hat{\rho}\hat{B})
    = \bigl(\hat{B}^\mathsf T \otimes \hat A\bigr)\,\mathrm{vec}(\hat{\rho}),
    \label{eq:vec_identity}
\end{equation}
valid for any operators $\hat A$ and $\hat{B}$, \cref{eq:ME_total_AG} can be written as
\begin{equation}
    \frac{d\mathrm{vec}(\hat{\rho})}{dt} = \mathbf{L}(t)\,\mathrm{vec}(\hat{\rho}) = \left(\mathbf{L}_H(t)\ + \mathbf{L}_\mathrm{diss}\right)\,\mathrm{vec}(\hat{\rho}),
    \label{eq:vec_ode}
\end{equation}
where $\mathbf{L}(t)$ is the total $d^2\times d^2$ Liouvillian matrix. The term $-\mathrm{i}\bigl[\hat H_{\mathrm{eff}}(t),\hat{\rho}\bigr]$ with
$\hat H_{\mathrm{eff}}(t) = \hat H_\mathrm{s} + \hat H_\mathrm{LS} + \hat H_\mathrm{drive}(t)$
yields
\begin{equation}
    \mathbf{L}_H(t) = -\mathrm{i}\Bigl(\mathbb I_d \otimes \hat H_{\mathrm{eff}}(t) - \hat H_{\mathrm{eff}}^\mathsf T(t) \otimes \mathbb I_d\Bigr),
\label{eq:LH}
\end{equation}
and $\mathbf{L}_\mathrm{diss} = \sum_i\mathbf{L}_{\hat{\mathcal{J}}_i}$ is the matrix representation of the total dissipator $\mathcal{L}_\mathrm{tot}$ in \cref{eq:Ltotal}, where each jump operator $\hat{\mathcal{J}}_i \in \{\hat{Q}_\mu, \hat{L}_\uparrow, \hat{L}_\downarrow, \hat{N}_{+}, \hat{N}_{-}\}$ contributes to the $d^2 \times d^2$ matrix representation of the superoperator $\mathcal{D}[\hat{\mathcal{J}}_i]$,
\begin{equation}
    \mathbf{L}_{\hat{\mathcal{J}}_i}
    = \hat{\mathcal{J}}_i^* \otimes \hat{\mathcal{J}}_i
    - \frac{1}{2}\,\mathbb I_d \otimes (\hat{\mathcal{J}}_i^\dagger \hat{\mathcal{J}}_i)
    - \frac{1}{2}\,(\hat{\mathcal{J}}_i^\dagger \hat{\mathcal{J}}_i)^\mathsf T \otimes \mathbb I_d.
    \label{eq:LL}
\end{equation}
\cref{eq:vec_ode,eq:LH,eq:LL} define a closed system of $d^2$ coupled first-order linear differential equations for the components of $\mathrm{vec}(\hat{\rho})$, which can be integrated directly using standard numerical methods.

\section{The dynamics of bright and dark polaritons}
\label{s:secular_breakdown}

The QME~\eqref{eq:ME_total_AG} provides a natural starting point for comparing the leakage dynamics predicted by the full, non-secular description and its secular approximation. It is especially illuminating to reorganize the UP/LP polariton branches into \emph{bright} and \emph{dark} superpositions, which clarify the structure of the system–bath coupling.

In the numerical calculations, we use the truncated polariton Fock space with $n_+, n_- \le 1$ (dimension $d=4$), spanned by $\{\ket{0,0}, \ket{0,1}, \ket{1,0}, \ket{1,1}\}$, where $\ket{n_+, n_-}$ denotes the occupation of the upper and lower polariton modes. The bright and dark single-excitation states are defined in the one-excitation subspace $\mathrm{span}\{\ket{0,1}, \ket{1,0}\}$ as
\begin{equation}
    \ket{\mathrm{b}} = \frac{\ket{1,0} + \ket{0,1}}{\sqrt{2}},
    \qquad
    \ket{\mathrm{d}} = \frac{\ket{1,0} - \ket{0,1}}{\sqrt{2}} ,
\label{eq:bright_dark}
\end{equation}
and they are annihilated by the corresponding mode operators
\begin{equation}
    \hat{B}_{\mathrm{b}} = \frac{\hat{B}_+ + \hat{B}_-}{\sqrt{2}},
    \qquad
    \hat{B}_{\mathrm{d}} = \frac{\hat{B}_+ - \hat{B}_-}{\sqrt{2}}.
    \label{eq:B_bright_dark_BR}
\end{equation}
Because the truncation includes the double-excitation state $\ket{1,1}$, the number operators $\hat{N}_{\mathrm{b},\mathrm{d}}=\hat{B}_{\mathrm{b},\mathrm{d}}^\dagger\hat{B}_{\mathrm{b},\mathrm{d}}$ receive contributions from both the single-excitation manifold and the $\ket{1,1}$ sector. To quantify the bright and dark populations, we therefore use the projectors
\begin{equation}
    \hat{P}_{\mathrm{b}}=\ket{\mathrm{b}}\!\bra{\mathrm{b}},
    \qquad
    \hat{P}_{\mathrm{d}}=\ket{\mathrm{d}}\!\bra{\mathrm{d}},
\label{eq:projectors_bd}
\end{equation}
whose expectation values $\langle \hat{P}_{\mathrm{b},\mathrm{d}}(t)\rangle=\mathrm{Tr}[\hat{P}_{\mathrm{b},\mathrm{d}}\hat{\rho}(t)]$ are dimensionless probabilities in $[0,1]$. We also monitor the UP-LP coherence in the single-excitation manifold, 
\begin{equation}
    |\rho_\mathrm{c}(t)|=|\langle 1,0|\hat{\rho}(t)|0,1\rangle| ,
\end{equation}
since the non-secular damping cross-terms act precisely through these coherences.
The transition operators
\begin{align}
    \hat{\sigma}_{1}\equiv \hat{B}_{+} &= \ket{0,0}\bra{1,0}+\ket{0,1}\bra{1,1},
    \nonumber \\
    \hat{\sigma}_{2}\equiv \hat{B}_{-} &= \ket{0,0}\bra{0,1}+\ket{1,0}\bra{1,1},
\end{align}
collect the two degenerate lowering transitions at Bohr frequencies $\omega_{1}\simeq \omega_{+}$ and $\omega_{2}\simeq \omega_{-}$, respectively (strict equality is usually assumed, neglecting the small shifts due to coherent mixing and bath-induced renormalization when evaluating the rates). In contrast, the bright and dark single-excitation states $\ket{\mathrm{b}}$ and $\ket{\mathrm{d}}$ are superpositions of eigenstates with different Bohr frequencies; consequently, the transitions from $\ket{\mathrm{b}}$ and $\ket{\mathrm{d}}$ to the vacuum are not associated with a unique frequency, but involve both $\omega_{+}$ and $\omega_{-}$.

\subsection{Breakdown of the secular approximation}

According to \cref{eq:aL_gG_means,eq:gamma}, the diagonal elements $W_{11}(\omega_1)$ and $W_{22}(\omega_2)$ are given by $J(\omega_1)$ and $J(\omega_2)$, respectively, while the off-diagonal effective dissipative terms are
\begin{equation}
    \widetilde{W}_{12}
    = \sqrt{W_{12}(\omega_1)\,W_{12}(\omega_2)}
    = \sqrt{J(\omega_1)\,J(\omega_2)}.
    \label{eq:G12_geo}
\end{equation}
For a Lorentzian $J(\omega)$ of width $\gamma_\mathrm{D}$ centered near the plasmonic resonance, this expression yields two distinct regimes:
\begin{itemize}
    \item If $|\Delta|\lesssim \gamma_\mathrm{D}/2$, then $J(\omega_1)\simeq J(\omega_2)$ and Eq.~\eqref{eq:G12_geo} implies $\widetilde W_{12}\sim W_{11}\sim W_{22}$. In this unresolved regime, the two transitions are effectively coupled by the same reservoir, so cross-damping and interference cannot be neglected: the dissipation eigenchannels become collective bright and dark modes, the bath-induced coherence is non-negligible, and long-lived dark mode may emerge. Therefore the full non-secular leakage kernel in $\mathcal{L}_\mathrm{tot}$ must be retained. In the numerical implementation, for the two leakage channels $\hat{\sigma}_1\equiv\hat{B}_+$ and $\hat{\sigma}_2\equiv\hat{B}_-$, the geometrically averaged dissipative kernel entering the non-secular leakage term is taken as 
    \footnote{for a Lorentzian spectrum of FWHM $\gamma_\mathrm{D}$ the bath correlation function decays as $C(\tau)\propto e^{-(\gamma_\mathrm{D}/2)\tau}$ (up to a carrier oscillation at $\omega_\mathrm{pl}$), and the oscillatory cross term carrying the beat phase $e^{\mathrm{i}\Delta\tau}$ is reduced by the normalized overlap
    \begin{equation}
    \kappa_\mathrm{c}
    =\frac{\int_0^\infty d\tau\,e^{-(\gamma_\mathrm{D}/2)\tau}\cos(\Delta\tau)}
    {\int_0^\infty d\tau\,e^{-(\gamma_\mathrm{D}/2)\tau}}
    =\frac{(\gamma_\mathrm{D}/2)^2}{(\gamma_\mathrm{D}/2)^2+\Delta^2} .
    \end{equation}}
    \begin{align}
        \widetilde{W} &=
        \begin{pmatrix}
        J(\omega_+) & g \\[2pt]
        g & J(\omega_-)
        \end{pmatrix}, \nonumber \\
        \kappa_\mathrm{c} &= \frac{(\gamma_\mathrm{D}/2)^2}{(\gamma_\mathrm{D}/2)^2+\Delta^2},
        \qquad
        g=\kappa_\mathrm{c}\sqrt{J(\omega_+)J(\omega_-)} ,
        \label{eq:Wleak_explicit}
    \end{align}
    where $\kappa_\mathrm{c}$ quantifies the suppression of cross-damping by finite bath memory, so that $\kappa_\mathrm{c}\to 1$ for $\Delta\ll\gamma_\mathrm{D}/2$ (unresolved transitions) and $\kappa_\mathrm{c}\to 0$ for $\Delta\gg\gamma_\mathrm{D}/2$.

    We call this \emph{common-bath} regime, because the bath correlation time $\/\gamma_\mathrm{D}$ is long enough that the bath cannot distinguish the $\omega_\pm$ transitions, so they effectively couple to the same reservoir mode continuum and acquire appreciable cross damping $g$.

    The Hermitian matrix $\widetilde{W}$ is then diagonalized as $\widetilde{W}=U\,\mathrm{diag}(d_\mu)\,U^\dagger$ and, if needed, projected onto its positive part by setting $d_\mu^{(+)}=\max(d_\mu,0)$, where 
    \begin{align}
    d_{\pm}&=\frac{J(\omega_+)+J(\omega_-)}{2} \nonumber \\
    &\pm \sqrt{\left(\frac{J(\omega_+)-J(\omega_-)}{2}\right)^2+g^2}\, .
    \label{eq:dpm_eigs}
    \end{align}
    This yields a GKSL-form dissipator with non-negative rates. For the $2\times 2$ kernel in Eq.~\eqref{eq:Wleak_explicit} one may write $\widetilde{W}=U\,\mathrm{diag}(d_{+},d_{-})\,U^\dagger$ with a rotation $U(\theta)$ defined by $\tan(2\theta)=2g/[J(\omega_+)-J(\omega_-)]$, so that the jump operators are linear combinations of the original transitions, $\hat{Q}_{+}=\cos\theta\,\hat{\sigma}_1+\sin\theta\,\hat{\sigma}_2$ and $\hat{Q}_{-}=-\sin\theta\,\hat{\sigma}_1+\cos\theta\,\hat{\sigma}_2$. Thus, the off-diagonal coefficient $g=\widetilde{W}_{12}$ is absorbed into the structure of $\hat{Q}_\pm$ (and into the eigenvalues $d_\pm$), rather than being discarded. The resulting generator is of GKSL form and therefore defines completely positive and trace-preserving reduced dynamics. 

    In the symmetric case $J(\omega_+)=J(\omega_-)\equiv\Gamma$, one has $d_{\pm}=\Gamma\pm g=\Gamma(1\pm\kappa_\mathrm{c})$, and the jump operators reduce to the bright and dark superpositions $\hat Q_{\pm}\propto \hat B_{+}\pm \hat B_{-}$ (Hadamard rotation).
    \item If $|\Delta|\gg\gamma_\mathrm{D}$, the bath resolves the two transition frequencies on its correlation-time scale and the oscillatory cross terms average out. Accordingly, the off-diagonal coefficients are strongly suppressed, $\widetilde W_{12}\ll W_{11},W_{22}$, and the leakage is well described by independent decay channels governed by $J(\omega_1)$ and $J(\omega_2)$. In this resolved regime the secular approximation is justified and the secular and non-secular generators yield practically identical dynamics \footnote{For a single-peaked spectral density, e.g., a Lorentzian $J(\omega) = J_0 \omega / [(\omega - \omega_0)^2 + \gamma_\mathrm{D}^2]$, taking two transition frequencies $\omega_i = \omega_0$ and $\omega_j = \omega_0 + \Delta$ gives $\sqrt{J(\omega_i) J(\omega_j)} = J_0 [1 + (\Delta / \gamma_\mathrm{D})^2]^{-1/2}$, which is suppressed by a factor approximately equal to $\gamma_\mathrm{D} / |\Delta|$ when $|\Delta| \gg \gamma_\mathrm{D}$.}. 
    The leakage dynamics are well described by a dissipator of the form
    \begin{equation}
    \mathcal{L}_\mathrm{leak}^{(\mathrm{sec})}\hat{\rho} \simeq     \sum_{k=\pm}\Gamma_k \left(\hat{B}_k\hat{\rho}\hat{B}_k^\dagger - \tfrac{1}{2}\bigl\{\hat{B}_k^\dagger\hat{B}_k, \hat{\rho}\bigr\} \right),
    \label{eq:L_secular_final}
    \end{equation}
    with decay rates $\Gamma_k$ obtained from $J(\omega_k)$ via the microscopic self-energy $\mathcal{S}(\omega_k)$. It is important to emphasize that the QME \cref{eq:ME_total_AG,eq:Ltotal} retains the non-secular couplings while guaranteeing complete positivity, whereas a fully secular Lindblad equation based on independent decay channels often predicts much faster decay of dark excitations and substantially different steady-state populations.
\end{itemize}

In the present approach, the separation $\Delta$ is controlled by the light–matter coupling strength $\xi\Omega_\mathrm{pl}$ and by the losses $\gamma_\mathrm{D}$ as described in Sec.~\ref{s:model} and in the Appendix \ref{app:eigenfrequencies}, \cref{eq:d2}. Our microscopic construction thus establishes a direct connection between the internal strong- or ultrastrong-coupling physics (which sets $\Delta$) and the validity of the secular approximation for the leakage dynamics (which depends on $\Delta/\gamma_\mathrm{D}$). Even in the weak light–matter coupling regime, non-secular effects can be significant if $\Delta$ is comparable to the bath linewidth, while in the ultrastrong-coupling regime, the increased separation can push the system toward the secular limit, provided that $\Delta$ exceeds the Drude–Lorentz width.

A distinctive feature of the common-bath regime is the potential emergence of an approximately decoherence-free dark polariton, as confirmed by experimental results \cite{2008Chong_PRA,2023Kim_PRL,2025Bouteyre_OEX}. Specifically, when $\Delta \simeq \gamma_\mathrm{D}$, the Kossakowski matrix associated with the leakage channel has a bright eigenmode proportional to $(\sigma_+ + \sigma_-)$ with a finite eigenvalue, and an orthogonal dark eigenmode proportional to $(\sigma_+ - \sigma_-)$ with a numerically vanishing eigenvalue. Consequently, the corresponding jump operators act only on the bright superposition $\hat{B}_\mathrm{b}$, while the dark superposition $\hat{B}_\mathrm{d}$ does not couple directly to the common bath and remains strictly decoherence-free at this level of approximation. When incoherent UP$\leftrightarrow$LP scattering and pure dephasing channels are included, this exact protection is lost: these additional processes do not preserve the bright/dark decomposition and introduce leakage pathways for $\hat{B}_\mathrm{d}$, turning the dark polariton into a \emph{quasi}-protected mode with a finite yet parametrically longer lifetime than the bright one.

\subsection{Numerical comparison of secular and non-secular dynamics}

To explicitly compare the dynamics obtained from the full non-secular master equation and its secular approximation, we numerically integrate the time-local GKSL master equation in the considered truncated polariton Fock space, so the Liouvillian superoperator acts on a $d^2 = 16$-dimensional vectorized density matrix. A convenient ordered basis is $|1\rangle = |0,0\rangle,\,
    |2\rangle = |0,1\rangle,\,
    |3\rangle = |1,0\rangle,\,
    |4\rangle = |1,1\rangle$. The vectorized form of the density matrix is
\begin{align}
    \mathrm{vec}(\hat{\rho})
    &=
    \begin{pmatrix}
    \rho_{11},\ \rho_{21},\ \rho_{31},\ \rho_{41},\ 
    \rho_{12},\ \dots,\ \rho_{44}
    \end{pmatrix}^{\!\mathsf T},
    \label{eq:rho_vec_N1}
\end{align}
where the columns of $\hat{\rho}$ are stacked into a single column vector of length $d^2=16$.

The single-mode annihilation operator in the truncated basis $\{|0\rangle,|1\rangle\}$ is
\begin{equation}
    \hat{b} =
    \begin{pmatrix}
        0 & 1 \\
        0 & 0
    \end{pmatrix},
    \label{eq:b_truncated}
\end{equation}
and the polariton annihilation operators $\hat{B}_{\pm}$ are built as
\begin{equation}
    \hat{B}_+ = \hat{b}\otimes \mathbb{I}_2,
    \qquad
    \hat{B}_- = \mathbb{I}_2\otimes \hat{b},
    \label{eq:B_tensor}
\end{equation}
which yield explicitly
\begin{equation}
    \hat{B}_+ =
    \begin{pmatrix}
        0 & 0 & 1 & 0 \\
        0 & 0 & 0 & 1 \\
        0 & 0 & 0 & 0 \\
        0 & 0 & 0 & 0
    \end{pmatrix},
    \qquad
    \hat{B}_- =
    \begin{pmatrix}
        0 & 1 & 0 & 0 \\
        0 & 0 & 0 & 0 \\
        0 & 0 & 0 & 1 \\
        0 & 0 & 0 & 0
    \end{pmatrix}.
    \label{eq:Bpm_matrices}
\end{equation}

We consider a two-stage driving protocol that prepares bright and dark superpositions and then couples them coherently. Starting from an initial vacuum state $\hat{\rho}(0)=\ket{0,0}\!\bra{0,0}$, a coherent drive with amplitude $f$ and frequency $\omega_\mathrm{c}$ is switched on at $t=0$ and illuminates the cavity for $0 \le t < t_\mathrm{sw}$ (stage-1), which populates the symmetric (bright) superposition of upper and lower polaritons,
\begin{equation}
    \hat{H}^{(\mathrm{b})}_{\mathrm{drive}}
    = f\left(\hat{B}_+ + \hat{B}_- + \hat{B}_+^\dagger + \hat{B}_-^\dagger\right),
    \qquad 0 \le t < t_\mathrm{sw}.
    \label{eq:H_drive_bright}
\end{equation}
In this regime the Raman-like UP-LP coupling is switched off. At $t=t_\mathrm{sw}$  (stage-2) we turn on a coherent interbranch (Raman-like) coupling of strength $\mathcal{R}_{\mathrm{UPLP}}$,
\begin{equation}
    \hat{H}_\mathrm{R}
    = \mathcal{R}_{\mathrm{UPLP}}
    \bigl(\hat{B}_+^\dagger\hat{B}_- + \hat{B}_-^\dagger\hat{B}_+\bigr),
    \qquad t_\mathrm{sw} \le t  <  t_\mathrm{off},
    \label{eq:H_Raman_num}
\end{equation}
and change the illumination pattern to a \emph{dark} drive that populates the antisymmetric superposition,
\begin{equation}
    \hat{H}_{\mathrm{drive}}^{(\mathrm{d})}
    = f\left(\hat{B}_+ - \hat{B}_- + \hat{B}_+^\dagger - \hat{B}_-^\dagger\right) .
\label{eq:H_drive_dark}
\end{equation}
For $t \geq t_\mathrm{off}$, the coherent drives and the Raman coupling are switched off, and the system evolves freely under leakage, scattering, and dephasing.

While \cref{eq:vec_ode,eq:LH,eq:LL} provide a compact vectorized form useful for discussion, for numerical integration it is necessary to write the corresponding system of coupled first-order ODEs for the density-matrix elements explicitly in the truncated polariton Fock space $n_\pm \leq 1$. The derivation is given in Appendix\,\ref{app:component_odes}, which provides the time evolution of the bright and dark projectors, $\langle \hat{P}_{\mathrm{b}}(t) \rangle$ and $\langle \hat{P}_{\mathrm{d}}(t) \rangle$.
\begin{figure*}[!t]
\centerline{
\includegraphics[width=1\textwidth]{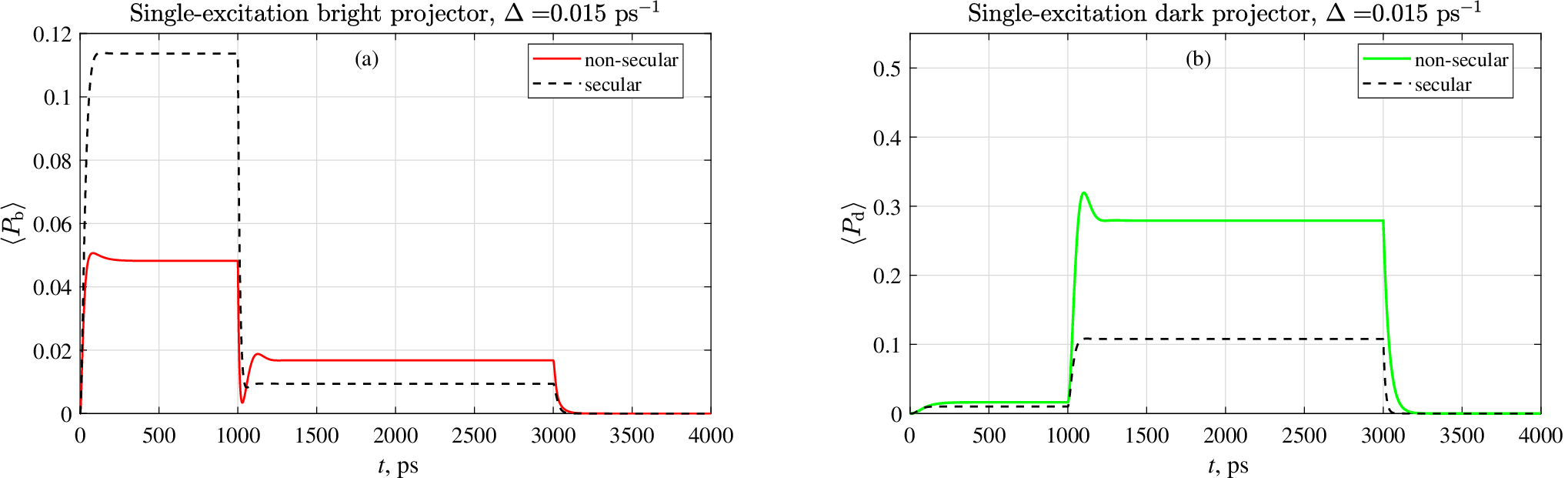}}
\vspace{0.5cm}
\centerline{
\includegraphics[width=1\textwidth]{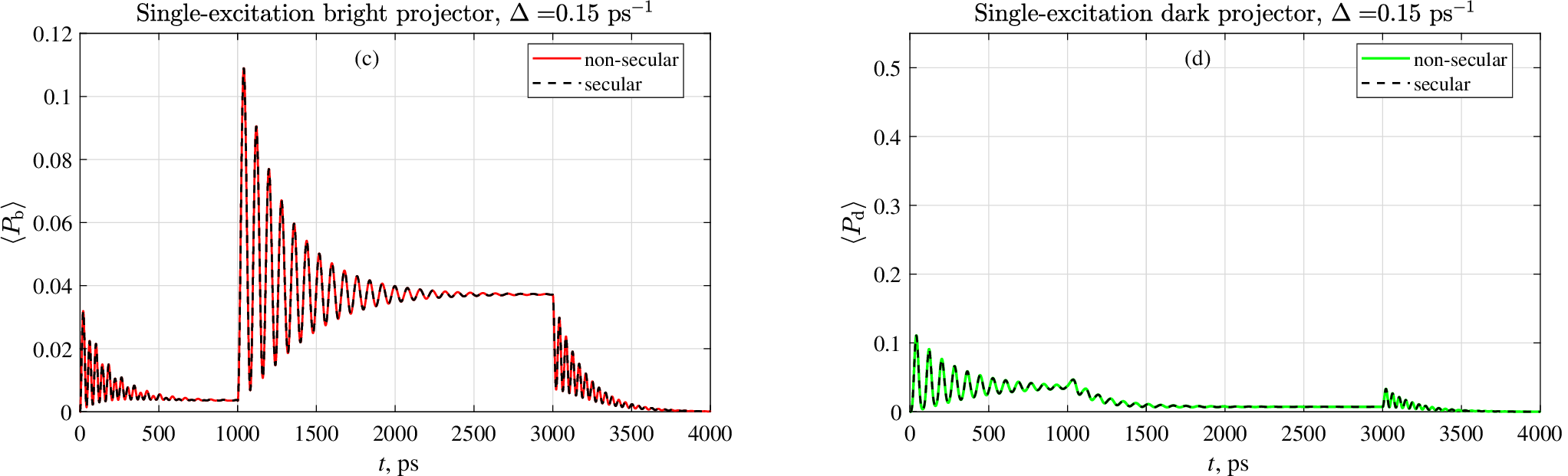}}
\caption{Time evolution under the two-stage drive protocol of the single-excitation bright and dark populations, $\langle \hat{P}_{\mathrm{b}}(t)\rangle$ and $\langle \hat{P}_{\mathrm{d}}(t)\rangle$. Solid lines: full non-secular common-bath leakage; dashed lines: secular approximation (diagonal Kossakowski matrix). Parameters: $T=300$\,K, $\gamma_\mathrm{D}=0.04$\,ps$^{-1}$, $\Gamma_{\mathrm{pl}}=2\gamma_\mathrm{D}$, $\gamma_{\downarrow}=10^{-3}$\,ps$^{-1}$ (with $\gamma_{\uparrow}$ fixed by detailed balance), $\gamma_{\phi}=5\times10^{-4}$\,ps$^{-1}$, and drive amplitude $f=0.01$\,ps$^{-1}$. The bright drive is applied for $0\le t<t_\mathrm{sw}$ with $t_\mathrm{sw}=1000$\,ps, while for $t_\mathrm{sw}\le t<t_\mathrm{off}$ (with $t_\mathrm{off}=3000$\,ps) the illumination is switched to the dark drive and a Raman-like coupling $\mathcal{R}_\mathrm{UP\!-\!LP}=0.01$\,ps$^{-1}$ is turned on. Panels (a,b): $\Delta=0.015$\,ps$^{-1}$; panels (c,d): $\Delta=0.15$\,ps$^{-1}$.}
\label{f:fig_2}
\end{figure*}

To highlight the role of the UP-LP frequency separation $\Delta$ relative to the bath linewidth, in Fig.\,\ref{f:fig_2} we compare the results obtained for two cases: a regime with $\Delta = 0.015$\,ps$^{-1} \lesssim \gamma_\mathrm{D}$ and a well-resolved regime $\Delta = 0.15$\,ps$^{-1} \gg \gamma_\mathrm{D}$. The numerical parameters used are: bath linewidth (correlation rate) $\gamma_\mathrm{D} = 0.04$\,ps$^{-1}$, central frequency $\omega_\mathrm{c} = 1$\,ps$^{-1}$, temperature $T = 300$\,K, scattering rate $\gamma_{\downarrow} = 10^{-3}$\,ps$^{-1}$ with $\gamma_{\uparrow}$ fixed by detailed balance, and pure dephasing $\gamma_{\phi} = 5 \times 10^{-4}$\,ps$^{-1}$. The coherent drive amplitude $f$ and the Raman-like coupling strength $\mathcal{R}_{\mathrm{UPLP}}$ are $0.01$\,ps$^{-1}$; we take $t_\mathrm{sw} = 1000$\,ps, $t_\mathrm{off}=3000$\,ps, and integrate up to $t_\mathrm{max}=4000$\,ps.

In the first case (Fig.\,\ref{f:fig_2}(a, b)), the leakage bath cannot resolve the UP and LP frequencies, so the geometric-mean cross rates in the Kossakowski matrix $\widetilde{W}_{12}$ remain significant, and the non-secular dynamics produces visible deviations from the secular approximation, most clearly in the build-up and decay of the dark-state population. 
In contrast, for $\Delta \gg \gamma_\mathrm{D}$ (Fig.\,\ref{f:fig_2}(c, d)), the secular and non-secular leakage models become practically indistinguishable throughout the driven dynamics, due to the suppression of the effective cross rates $\widetilde{W}_{12}$.

Approximate steady-state solutions for $\langle \hat P_\mathrm{b,d}\rangle$ can be obtained by setting the derivative in \cref{eq:vec_ode,eq:LH,eq:LL} to zero. The procedure is described in detail in Appendix\,\ref{app:steadystate}, neglecting the Raman-like coupling and UP-LP scattering.
In the secular approximation, this yields the following expressions ("ss" for steady state and "sec" for secular):
\begin{align}
\langle \hat P_{\mathrm b}\rangle_{\mathrm ss}^{(\mathrm{sec})}\big|_{\text{bright drive}}
&=\frac{4 f^{2}}{D^{2}}\left(2\Gamma^{2}+4 f^{2}\right) ,\nonumber\\
\langle \hat P_{\mathrm d}\rangle_{\mathrm ss}^{(\mathrm{sec})}\big|_{\text{bright drive}}
&=\frac{4 f^{2}}{D^{2}}\left(2\Delta^{2}+4 f^{2}\right) ,
\label{eq:PbPd_sec_bright}
\end{align}
for the bright drive, while for the dark drive the two expressions are exchanged,
\begin{align}
\langle \hat P_{\mathrm b}\rangle_{\mathrm ss}^{(\mathrm{sec})}\big|_{\text{dark drive}}
&=\langle \hat P_{\mathrm d}\rangle_{\mathrm ss}^{(\mathrm{sec})}\big|_{\text{bright drive}} \nonumber \\
\langle \hat P_{\mathrm d}\rangle_{\mathrm ss}^{(\mathrm{sec})}\big|_{\text{dark drive}}
&=\langle \hat P_{\mathrm b}\rangle_{\mathrm ss}^{(\mathrm{sec})}\big|_{\text{bright drive}}.
\label{eq:PbPd_sec_dark}
\end{align}
where $\Gamma$ is given by \cref{eq:Gamma_pm_def}, and $D = \Gamma^{2}+\Delta^{2}+8 f^{2}$.

\begin{figure*}[!t]
\centering
\includegraphics[width=1\textwidth]{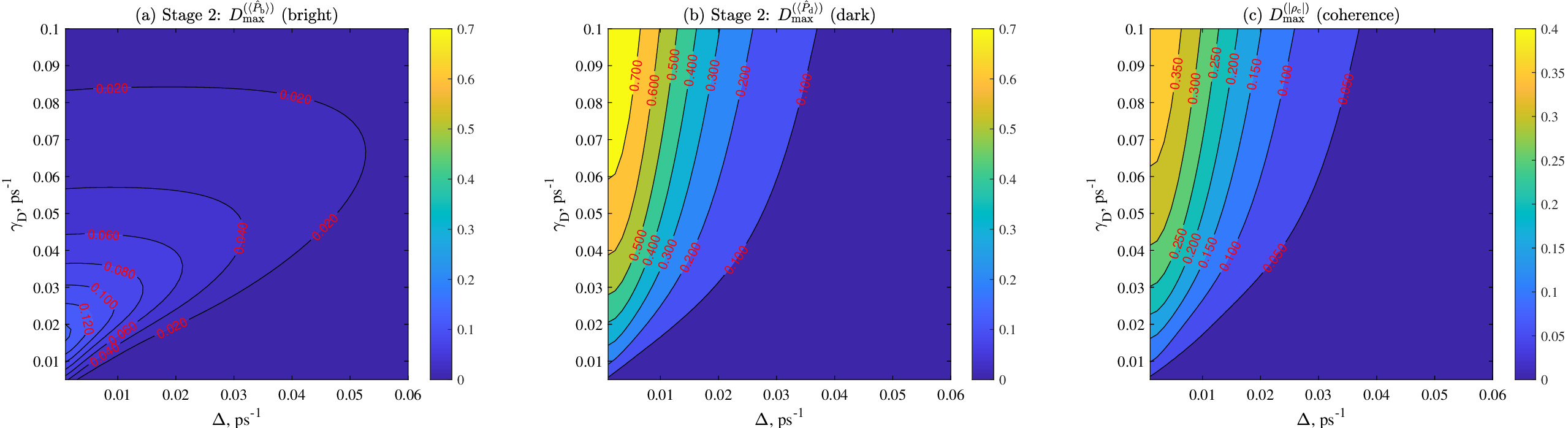}
\caption{(Color online) Stage-2 maximum deviation $D_{\max}^{(X)}(\Delta,\gamma_\mathrm{D})$, \cref{eq:Dmax_def}, between the non-secular common-bath leakage model and its secular approximation, for $X=\langle\hat{P}_\mathrm{b}\rangle$ (a), $\langle\hat{P}_\mathrm{d}\rangle$ (b), and $|\rho_\mathrm{c}|$ (c), evaluated over the stage-2 time window of Fig.~\ref{f:fig_2} while scanning $\Delta$ and $\gamma_\mathrm{D}$. The bath spectrum is Lorentzian, \cref{eq:J_Lorentz}, with FWHM $\gamma_\mathrm{D}$ and peak rate $\Gamma_\mathrm{pl}\equiv J(\omega_\mathrm{pl})=2\gamma_\mathrm{D}$. Weak coherent couplings: $f=\mathcal{R}_\mathrm{UP\!-\!LP}=0.01~\mathrm{ps}^{-1}$.}
\label{f:fig_3}
\end{figure*}
\begin{figure*}[!t]
\centering
\includegraphics[width=1\textwidth]{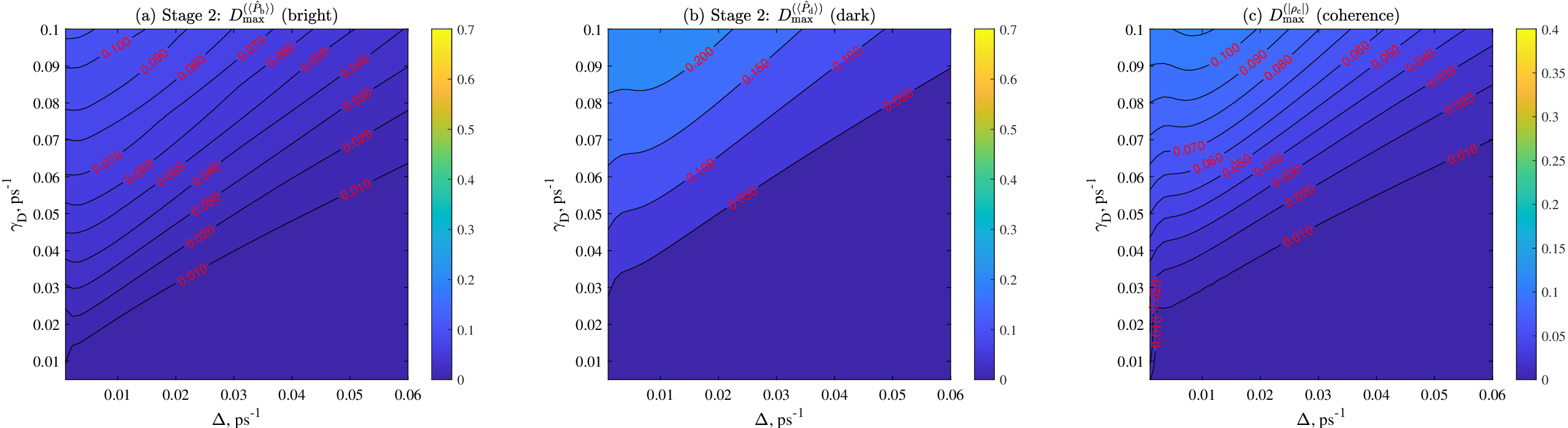}
\caption{(Color online) Same as Fig.~\ref{f:fig_3}, but for stronger coherent couplings $f=\mathcal{R}_\mathrm{UP\!-\!LP}=0.1~\mathrm{ps}^{-1}$.}
\label{f:fig_4}
\end{figure*}

When the non-secular terms are retained, the presence of the cross-damping terms need a different approach, also described in Appendix\,\ref{app:steadystate}. For the two drives, the steady-state solutions for $\langle \hat P_\mathrm{b,d}\rangle$ are ("ns" standing for non-secular)
\begin{align}
\label{eq:ss_b_drive_ns}
\langle \hat P_{\mathrm b}\rangle_{\mathrm ss}^{(\mathrm{ns})}\big|_{\text{bright drive}} &=\frac{32 f^{2}\Bigl[\Gamma^2 + 2 f^{2}\Bigr]}{\mathcal N_{\mathrm b}} , \nonumber \\
\langle \hat P_{\mathrm d}\rangle_{\mathrm ss}^{(\mathrm{ns})}\big|_{\text{bright drive}} &=\frac{64 f^{4}}{\mathcal N_{\mathrm b}}  
\end{align}
\begin{align}
\label{eq:ss_d_drive_ns}
\langle \hat P_{\mathrm b}\rangle_{\mathrm ss}^{(\mathrm{ns})}\big|_{\text{dark drive}} &=\frac{64 f^{4}}{\mathcal N_{\mathrm b}} , \nonumber \\
\langle \hat P_{\mathrm d}\rangle_{\mathrm ss}^{(\mathrm{ns})}\big|_{\text{dark drive}} &=\frac{32 f^{2}\Bigl[\Gamma^2 + 2 f^{2}\Bigr]}{\mathcal N_{\mathrm d}} ,
\end{align}
where $\mathcal N_\mathrm{b,d} = 4 \Gamma^2 \bigl(\Gamma_\mathrm{b,d}^{2}+16 f^{2}\bigr) + (4f)^4$.

With parameters as in Fig.\,\ref{f:fig_2} and $\Delta = 0.015$, the steady state values provided by \cref{eq:PbPd_sec_bright,eq:PbPd_sec_dark,eq:ss_b_drive_ns,eq:ss_d_drive_ns} are shown in Table\,\ref{tab:table1}, which are in good agreement with Fig.\,\ref{f:fig_2}.
\begin{table}[h]
    \centering
    \caption{stationary solutions, $\Delta = 0.015\,\text{ps}^{-1}$}
    \label{tab:placeholder_label}
    \begin{tabular}{ccccc}
        \toprule
                         & $\langle\hat P_{\mathrm b}\rangle_{\mathrm ss}^{(\mathrm{ns})}$
                         & $\langle\hat P_{\mathrm b}\rangle_{\mathrm ss}^{(\mathrm{sec})}$ 
                         & $\langle\hat P_{\mathrm d}\rangle_{\mathrm ss}^{(\mathrm{ns})}$ 
                         & $\langle\hat P_{\mathrm d}\rangle_{\mathrm ss}^{(\mathrm{sec})}$ \\
        \midrule
        bright drive & 0.056 & 0.116 & $2.17 \times 10^{-3}$ & $9.6 \times 10^{-3}$ \\
        dark drive & 0.014 & $9.6 \times 10^{-3}$ & 0.352 & 0.116 \\
        \bottomrule
    \end{tabular}
\label{tab:table1}
\end{table}

For $|\Delta| \gg \gamma_{\mathrm D}/2$ (so that $\kappa_\mathrm{c} \to 0$), the cross-correlations are suppressed ($g \to 0$), and the natural jump operators approach $\hat B_{\pm}$ instead. As described in detail in Appendix\,\ref{app:steadystate}, the bright and dark ladder solutions given by \cref{eq:ss_b_drive_ns,eq:ss_d_drive_ns} are no longer representative, whereas the component rate equations and the numerical GKSL integration results (e.g., Fig.\,\ref{f:fig_2}, as well as Fig.\,\ref{f:fig_3} and Fig.\,\ref{f:fig_4} described in Section\,\ref{s:param_space}) remain valid for arbitrary $\Delta$.

\subsection{Parameter space investigation}
\label{s:param_space}
To identify the parameter region where non-secular leakage terms are quantitatively relevant and to provide design guidance, we perform a two-dimensional scan over the polariton splitting $\Delta$ and the Drude–Lorentz linewidth $\gamma_\mathrm{D}$, comparing the non-secular common-bath dynamics with its secular approximation under the same two-stage protocol as in Fig.~\ref{f:fig_2}. Since the interference-driven signatures (bath-induced coherence and bright/dark selectivity) develop mainly during the second stage (dark drive plus Raman-like coupling), we quantify discrepancies by the stage-2 maximum deviation,
\begin{equation}
D_{\max}^{(X)}(\Delta,\gamma_\mathrm{D})
=\max_{t\in[t_{\mathrm{sw}},t_{\mathrm{off}}]}\big|X_{\rm ns}(t)-X_{\rm sec}(t)\big|,
\label{eq:Dmax_def}
\end{equation}
with $X\in\{\langle\hat{P}_\mathrm{b}\rangle,\langle\hat{P}_\mathrm{d}\rangle,|\rho_\mathrm{c}|\}$.
The results are shown as color maps in Fig.~\ref{f:fig_3} for weak coherent couplings and in Fig.~\ref{f:fig_4} for stronger driving, in the $(\Delta,\,\gamma_\mathrm{D})$ parameter space.

For both drive strength, the largest deviations occur in the ``unresolved'' regime $\Delta\lesssim\gamma_\mathrm{D}$, where the reservoir cannot distinguish the UP and LP transition frequencies on the bath-correlation scale and the cross-damping terms in the common-bath kernel are not averaged out. For $\Delta\gg\gamma_\mathrm{D}$ the discrepancies rapidly collapse, consistent with the suppression of off-diagonal dissipative couplings.

In the weak-drive case (Fig. \ref{f:fig_3}), two distinct features appear at small $\Delta$. First, a localized enhancement of $D_{\max}^{(\langle\hat{P}_\mathrm{b}\rangle)}$ at small $\Delta$ and small $\gamma_\mathrm{D}$ originates from the bright population prepared during stage 1: when the leakage rate is slow on the stage-2 time window, a residual bright component survives the switching and enters stage 2, so the non-secular correction to the effective bright decay channel yields a visible mismatch in $\langle\hat{P}_\mathrm{b}\rangle$. Second, for larger $\gamma_\mathrm{D}$ at the same small $\Delta$, dissipation acts on the stage-2 timescale and the dominant difference becomes the fate of the dark combination: the secular approximation predicts independent decay governed by $J(\omega_\pm)$, whereas the non-secular common-bath kernel approaches a nearly rank-one form and partially protects the orthogonal dark state through destructive interference. This produces the vertical high-deviation region in $D_{\max}^{(\langle\hat{P}_\mathrm{d}\rangle)}$ and the concomitant enhancement of $D_{\max}^{(|\rho_\mathrm{c}|)}$ at small $\Delta$.

In the strong-drive case (Fig. \ref{f:fig_4}), the crossover between secular and non-secular behavior is still determined by the ratio $\Delta/\gamma_\mathrm{D}$, but the deviations are spread over a broader upper-left wedge. Compared to the weak-drive scan, the stage-2 discrepancies are generally reduced because the stronger dark drive and Raman coupling transiently populate higher-lying states (including $|1,1\rangle$). This lowers the instantaneous weight of the single-excitation projectors during stage 2 and reduces their absolute mismatches. In this regime, $D_{\max}^{(\langle\hat{P}_\mathrm{d}\rangle)}$ remains the most sensitive indicator, while $D_{\max}^{(\langle\hat{P}_\mathrm{b}\rangle)}$ and $D_{\max}^{(|\rho_\mathrm{c}|)}$ follow similar contours with smaller amplitudes.

\section{Conclusions and outlook}
\label{s:conclusions}

We developed a Markovian master-equation framework for leakage in a hybrid plasmonic cavity beyond the secular approximation, based on the Drude–Lorentz response of the medium and the associated photon self-energy $\mathcal{S}(\omega)$. The resulting time-local generator preserves off-diagonal Bloch–Redfield couplings between transitions at different Bohr frequencies and is completely positive.

In the bright/dark basis, the physical meaning of the non-secular leakage channel is clear. For a single-peaked Drude–Lorentz spectral density with width $\gamma_\mathrm{D}$, cross-damping scales as $\sqrt{J(\omega_+)J(\omega_-)}$ and becomes significant when the polariton splitting satisfies $\Delta\lesssim\gamma_\mathrm{D}$. In this regime, the two decay paths interfere and bath-induced coherences are generated: leakage acts mainly on the bright combination, while the orthogonal dark combination is approximately protected by destructive interference. Incoherent UP$\leftrightarrow$LP scattering and pure dephasing remove this protection and turn the dark polariton into a quasi-long-lived mode.

Numerical simulations in the truncated polariton Fock space with $n_+, n_- \le 1$ make these regimes explicit. By monitoring the single-excitation bright and dark projectors and the UP–LP coherence under a two-stage bright/dark drive protocol, we find that non-secular leakage can sustain dark-state population and bath-induced coherence, whereas the secular approximation can overestimate dark decay and suppress coherence when $\Delta \sim \gamma_\mathrm{D}$. For $\Delta \gg \gamma_\mathrm{D}$, the non-secular generator smoothly approaches the secular limit, and the two descriptions become essentially indistinguishable.

This framework provides a practical route to connect microscopic material response, Green's-function electrodynamics, and completely positive open-system dynamics in plasmonic cavity QED. Natural extensions include larger Fock-space truncations, multiple structured reservoirs (e.g., separating radiative and absorptive channels), inclusion of output fields to access emission spectra and correlation functions, and benchmarks against explicitly non-Markovian treatments when bath memory becomes comparable to system timescales.

\begin{acknowledgments}
This work was supported in part by the European Union under two initiatives of the Italian National Recovery and Resilience Plan (NRRP) of NextGenerationEU: the partnership on Telecommunications of the Future (PE00000001 – program “RESTART”), and the National Centre for HPC, Big Data and Quantum Computing (CN00000013 – CUP E13C22000990001).
\end{acknowledgments}

\appendix

\section{Cavity eigenfrequencies}
\label{app:eigenfrequencies}

By equating the denominator of $D(\omega)$ with the product $(\omega-\omega_{+})(\omega+\omega^*_{+})(\omega-\omega_{-})(\omega+\omega^*_{-})$ and requiring that the poles have the form 
\begin{equation}
\omega_\pm = \omega_\mathrm{c} \pm \zeta + \Sigma - \mathrm{i}\Gamma/2,
\end{equation}
in \cite{2025Vallone_ARX} we obtained
\begin{align}
\zeta^2 &= \frac{\omega_\mathrm{c}^2}{2} + \frac{\tau^2}{4} -\frac{3}{32}\gamma_\mathrm{D}^2  \nonumber \\
&\quad - \frac{1}{8}\sqrt{16\,\omega_\mathrm{c}^4 - \gamma_\mathrm{D}^2\!\left(2\omega_\mathrm{c}^2+\tau^2+\frac{5}{16}\gamma_\mathrm{D}^2\right)} ,
\label{eq:d2}
\end{align}
\begin{align}
\Sigma &= -\omega_\mathrm{c} 
+ \frac{1}{2}\!\left[2\omega_\mathrm{c}^2+\tau^2-\frac{3}{8}\gamma_\mathrm{D}^2 \right. \nonumber \\
&\quad \left. +2\!\sqrt{\omega_\mathrm{c}^4 
- \frac{1}{16}\!\left[\left(2\omega_\mathrm{c}^2+\tau^2\right)\gamma_\mathrm{D}^2 
+ 5\gamma_\mathrm{D}^4\right]}\right]^{1/2} , \nonumber \\[2mm]
\Gamma &= \gamma_\mathrm{D}/2 ,
\label{eq:SigmaGamma}
\end{align}
where $\tau \equiv \xi\,\Omega_\mathrm{pl}$ is the light-matter (EC-SPP) coupling strength. The frequencies separation,
\begin{equation}
(\omega_{+} - \omega_{-})_{\omega_\mathrm{pl}=\omega_\mathrm{c}} = 2 \zeta,
\end{equation}
is set entirely by the parameter $\zeta$ (the common shift $\Sigma$ cancels), which depends on the losses. In the lossless limit, we recover $\zeta = \tau/2$.

\section{Component rate equations}
\label{app:component_odes}

For numerical integration we use the time-local GKSL master equation
\cref{eq:ME_total_AG,eq:Ltotal} in the truncated polariton Fock space $n_\pm\le 1$.
Although \cref{eq:vec_ode,eq:LH,eq:LL} provide a compact vectorized form,
for numerical simulations it is necessary to write the corresponding system of coupled first-order
ODEs for the density-matrix elements explicitly, in a form tailored to the bright and dark
projectors $\hat P_{\mathrm b},\hat P_{\mathrm d}$.

\paragraph*{Basis and operators.}
We work in the rotated basis $\{\ket{\mathrm g},\ket{\mathrm b},\ket{\mathrm d},\ket{\mathrm e}\}$, where
$\ket{\mathrm b}=(\ket{+}+\ket{-})/\sqrt2$ and $\ket{\mathrm d}=(\ket{+}-\ket{-})/\sqrt2$.
In this basis, $\langle \hat P_{\mathrm b}(t)\rangle=\rho_{\mathrm{bb}}(t)$ and
$\langle \hat P_{\mathrm d}(t)\rangle=\rho_{\mathrm{dd}}(t)$ with
$\hat P_{\mathrm b}=\ket{\mathrm b}\!\bra{\mathrm b}$ and $\hat P_{\mathrm d}=\ket{\mathrm d}\!\bra{\mathrm d}$.

The truncated lowering operators are

\begin{align}
\hat B_{\mathrm b} &= \ket{\mathrm g}\bra{\mathrm b}+\ket{\mathrm b}\bra{\mathrm e}, \nonumber\\
\hat B_{\mathrm d} &= \ket{\mathrm g}\bra{\mathrm d}-\ket{\mathrm d}\bra{\mathrm e},
\label{eq:Bbd_trunc_app}
\end{align}
so that $\hat B_{\mathrm b}^\dagger\hat B_{\mathrm b}=\ket{\mathrm b}\bra{\mathrm b}+\ket{\mathrm e}\bra{\mathrm e}$
and $\hat B_{\mathrm d}^\dagger\hat B_{\mathrm d}=\ket{\mathrm d}\bra{\mathrm d}+\ket{\mathrm e}\bra{\mathrm e}$.
In the symmetric rotating frame of the central drive frequency, the system Hamiltonian
$\hat H_\mathrm{s}$ takes the form
\begin{equation}
\hat H_\mathrm{s}=\frac{\Delta}{2}\bigl(\ket{\mathrm b}\bra{\mathrm d}+\ket{\mathrm d}\bra{\mathrm b}\bigr),
\label{eq:Hs_bd_app}
\end{equation}
where $\Delta=\omega_+-\omega_-$ as in the main text (including, if desired, the corresponding Lamb shifts we always omit).
The (piecewise) coherent drive alternates between the bright $\hat H_{\mathrm{drive}}^{(\mathrm b)}$ ad dark $\hat H_{\mathrm{drive}}^{(\mathrm d)}$ drives,   
\begin{align}
\hat H_{\mathrm{drive}}^{(\mathrm b)}&=\sqrt2\,f(t)\bigl(\hat B_{\mathrm b}+\hat B_{\mathrm b}^\dagger\bigr),\nonumber\\
\hat H_{\mathrm{drive}}^{(\mathrm d)}&=\sqrt2\,f(t)\bigl(\hat B_{\mathrm d}+\hat B_{\mathrm d}^\dagger\bigr),
\end{align}
with real $f(t)$ (piecewise constant in the numerical examples).

\paragraph*{Leakage dissipator in the symmetric case.}
To keep the comparison transparent, we specialize \cref{eq:Wleak_explicit} to the symmetric case
$J(\omega_{+})=J(\omega_{-})\equiv\Gamma$.
In the $\{\hat B_{+},\hat B_{-}\}$ basis, the corresponding $2\times2$ Kossakowski matrix becomes
\begin{equation}
\widetilde{W}=
\begin{pmatrix}
\Gamma & g\\[2pt]
g & \Gamma
\end{pmatrix},
\qquad
g=\kappa_\mathrm{c}\Gamma.
\label{eq:Wpm_block_app}
\end{equation}
The eigenvectors are the symmetric/antisymmetric combinations, yielding the jump operators
\begin{equation}
\hat Q_{\mathrm b}=\frac{\hat B_{+}+\hat B_{-}}{\sqrt2}\equiv \hat B_{\mathrm b},
\qquad
\hat Q_{\mathrm d}=\frac{\hat B_{+}-\hat B_{-}}{\sqrt2}\equiv \hat B_{\mathrm d},
\label{eq:Qbd_from_Bpm_app}
\end{equation}
with corresponding rates
\begin{equation}
\Gamma_{\mathrm b}=\Gamma+g=\Gamma(1+\kappa_\mathrm{c}),\qquad 
\Gamma_{\mathrm d}=\Gamma-g=\Gamma(1-\kappa_\mathrm{c}).
\label{eq:Gam_bd_app}
\end{equation}
Substituting these into \cref{eq:L_leak} gives
\begin{equation}
\mathcal L_{\mathrm{leak}}^{(\mathrm{ns})}\hat{\rho}
=\Gamma_{\mathrm b}\,\mathcal D[\hat B_{\mathrm b}]\hat{\rho}
+\Gamma_{\mathrm d}\,\mathcal D[\hat B_{\mathrm d}]\hat{\rho}.
\label{eq:L_ns_app}
\end{equation}
In the secular approximation the cross term is dropped ($g=0$), so that $\Gamma_{\mathrm b}=\Gamma_{\mathrm d}\equiv\Gamma$.
Additional channels (UP$\leftrightarrow$LP scattering and pure dephasing) can be included by adding the corresponding
$\gamma_k\mathcal D[\hat L_k]$ and $\gamma_\phi\mathcal D[\hat N_\pm]$ terms as in \cref{eq:Ltotal}, although here we focus on the leakage-only case.
Using \cref{eq:vec_identity}, the vectorized ODE \cref{eq:vec_ode} is equivalent to
\begin{align}
&\dot\rho_{mn}
=-\mathrm{i}\sum_k\bigl(H_{mk}\rho_{kn}-\rho_{mk}H_{kn}\bigr)  \nonumber \\
&+ \sum_j \Gamma_j\Bigl[(J_j\rho J_j^\dagger)_{mn}-\tfrac12(J_j^\dagger J_j\rho+\rho J_j^\dagger J_j)_{mn}\Bigr],
\label{eq:component_general_app}
\end{align}
where $m,n,k\in\{\mathrm g,\mathrm b,\mathrm d,\mathrm e\}$, $\hat H=\hat H_\mathrm{s}+\hat H_\mathrm{drive}$,
and $\{J_j\}=\{\hat B_{\mathrm b},\hat B_{\mathrm d}\}$ with rates $\{\Gamma_j\}=\{\Gamma_{\mathrm b},\Gamma_{\mathrm d}\}$. 
The resulting closed subsystem of equations for the independent matrix elements (the remaining elements follow from Hermiticity, $\rho_{mn}=\rho_{nm}^*$, and $\mathrm{Tr}\,\rho=1$) depends on the type of drive. For the bright drive it is:
\begin{align}
\dot\rho_{\mathrm{gg}}&=\Gamma_{\mathrm b}\rho_{\mathrm{bb}}+\Gamma_{\mathrm d}\rho_{\mathrm{dd}} 
+\sqrt2\,\mathrm{i}f\bigl(\rho_{\mathrm{gb}}-\rho_{\mathrm{bg}}\bigr), \label{eq:ode_gg_bdrive_app}\\
\dot\rho_{\mathrm{bb}}&=-\Gamma_{\mathrm b}\bigl(\rho_{\mathrm{bb}}-\rho_{\mathrm{ee}}\bigr)
+\mathrm{i}\frac{\Delta}{2}\bigl(\rho_{\mathrm{bd}}-\rho_{\mathrm{db}}\bigr)  \nonumber \\
&+\sqrt2\,\mathrm{i}f\bigl(\rho_{\mathrm{bg}}-\rho_{\mathrm{gb}}+\rho_{\mathrm{be}}-\rho_{\mathrm{eb}}\bigr),\\
\dot\rho_{\mathrm{dd}}&=-\Gamma_{\mathrm d}\bigl(\rho_{\mathrm{dd}}-\rho_{\mathrm{ee}}\bigr)
+\mathrm{i}\frac{\Delta}{2}\bigl(\rho_{\mathrm{db}}-\rho_{\mathrm{bd}}\bigr),\\
\dot\rho_{\mathrm{ee}}&=-(\Gamma_{\mathrm b}+\Gamma_{\mathrm d})\rho_{\mathrm{ee}}
+\sqrt2\,\mathrm{i}f\bigl(\rho_{\mathrm{eb}}-\rho_{\mathrm{be}}\bigr),\\
\dot\rho_{\mathrm{gb}}&=-\frac{\Gamma_{\mathrm b}}{2}\rho_{\mathrm{gb}}+\Gamma_{\mathrm b}\rho_{\mathrm{be}}
+\mathrm{i}\frac{\Delta}{2}\rho_{\mathrm{gd}}  \nonumber \\
&+\mathrm{i}\sqrt2 f\bigl(\rho_{\mathrm{gg}}+\rho_{\mathrm{ge}}-\rho_{\mathrm{bb}}\bigr),\\
\dot\rho_{\mathrm{gd}}&=-\frac{\Gamma_{\mathrm d}}{2}\rho_{\mathrm{gd}}-\Gamma_{\mathrm d}\rho_{\mathrm{de}}
+\mathrm{i}\frac{\Delta}{2}\rho_{\mathrm{gb}}
-\mathrm{i}\sqrt2 f\,\rho_{\mathrm{bd}},\\
\dot\rho_{\mathrm{ge}}&=-\frac{\Gamma_{\mathrm b}+\Gamma_{\mathrm d}}{2}\rho_{\mathrm{ge}}
+\mathrm{i}\sqrt2 f\bigl(\rho_{\mathrm{gb}}-\rho_{\mathrm{be}}\bigr),\\
\dot\rho_{\mathrm{bd}}&=-\frac{\Gamma_{\mathrm b}+\Gamma_{\mathrm d}}{2}\rho_{\mathrm{bd}}
+\mathrm{i}\frac{\Delta}{2}\bigl(\rho_{\mathrm{bb}}-\rho_{\mathrm{dd}}\bigr)  \nonumber \\
&-\mathrm{i}\sqrt2 f\bigl(\rho_{\mathrm{gd}}+\rho_{\mathrm{ed}}\bigr),\\
\dot\rho_{\mathrm{be}}&=-\left(\Gamma_{\mathrm b}+\frac{\Gamma_{\mathrm d}}{2}\right)\rho_{\mathrm{be}}
-\mathrm{i}\frac{\Delta}{2}\rho_{\mathrm{de}}  \nonumber \\
&-\mathrm{i}\sqrt2 f\bigl(\rho_{\mathrm{ge}}-\rho_{\mathrm{bb}}+\rho_{\mathrm{ee}}\bigr),\\
\dot\rho_{\mathrm{de}}&=-\left(\Gamma_{\mathrm d}+\frac{\Gamma_{\mathrm b}}{2}\right)\rho_{\mathrm{de}}
-\mathrm{i}\frac{\Delta}{2}\rho_{\mathrm{be}}
+\mathrm{i}\sqrt2 f\,\rho_{\mathrm{db}} ,
\label{eq:ode_de_bdrive_app}
\end{align}
and similarly for the dark drive. The solutions shown in Figs.~\ref{f:fig_2}--\ref{f:fig_4} are obtained by integrating the above ODEs (with the appropriate choice of $\Gamma_{\mathrm b},\Gamma_{\mathrm d}$ for the non-secular description, or $\Gamma_{\mathrm b}=\Gamma_{\mathrm d}$ for the secular approximation) and then evaluating $\rho_{\mathrm{bb}}(t)$ and $\rho_{\mathrm{dd}}(t)$.

\section{Steady-state solution in the truncated polariton space}
\label{app:steadystate}

We work in the truncated polariton Fock space $n_{\pm}\le 1$ (dimension $d=4$), spanned by
\begin{equation}
\{\ket{\mathrm g},\ket{-},\ket{+},\ket{\mathrm e}\}\equiv \{|0,0\rangle,|0,1\rangle,|1,0\rangle,|1,1\rangle\},
\end{equation}
with $\ket{\mathrm g}$ the vacuum and $\ket{\mathrm e}$ the double-excitation state. The bright and dark states
$\ket{\mathrm b},\ket{\mathrm d}$ and the projectors $\hat P_{\mathrm b},\hat P_{\mathrm d}$ are defined in the main text according to \cref{eq:bright_dark,eq:projectors_bd}. In the $\{\ket{+},\ket{-}\}$ basis,
\begin{align}
\langle \hat{P}_{\mathrm b}\rangle &= \tfrac12 \left(\rho_{++}+\rho_{--}+\rho_{+-}+\rho_{-+}\right),\nonumber\\
\langle \hat{P}_{\mathrm d}\rangle &= \tfrac12 \left(\rho_{++}+\rho_{--}-\rho_{+-}-\rho_{-+}\right),
\label{eq:PbPd_elements_app}
\end{align}
where $\rho_{mn}\equiv\bra{m}\hat\rho\ket{n}$ ($m,n\in\{\mathrm g,-,+,\mathrm e\}$).

In the $\{\ket{+},\ket{-}\}$ basis, the piecewise coherent drive used throughout this work can be written as
\begin{align}
\hat H_{\mathrm{drive}}^{(\eta)}&=f\Big(\hat B_{+}+\eta\,\hat B_{-}+\hat B^\dagger_{+}+\eta\,\hat B^\dagger_{-}\Big), \nonumber \\
\eta&=
\begin{cases}
+1 & \text{(bright drive)} \\
-1 & \text{(dark drive)} .
\end{cases}
\label{eq:Hdrive_eta_app}
\end{align}
Thus the branch-resolved drive amplitudes $f_\mu$ are $f_{+}=f$ and $f_{-}=\eta f$. For clarity, for the bright drive ($\eta=+1$) one has $f_{+}=f_{-}=f$, whereas for the dark drive ($\eta=-1$) $f_{+}=f$ and $f_{-}=-f$.

\subsection{Secular approximation}
In the secular approximation, the two polariton branches evolve independently and the steady state (ss) factorizes,
$\rho_{\mathrm{ss}}^{(\mathrm{sec})}=\rho_{\mathrm{ss}}^{(+)}\otimes\rho_{\mathrm{ss}}^{(-)}$.
For each branch $\mu\in\{+,-\}$, we truncate to $\mathcal H_{\mu}^{(2)}=\mathrm{span}\{\ket{\mathrm g},\ket{\mu}\}$ and define $\hat{\sigma}^{(\mu)}\equiv \ket{\mathrm g}\bra{\mu}$ and $\hat{\sigma}^{(\mu) \dagger}\equiv \ket{\mu}\bra{\mathrm g}$.
In the rotating frame of the drive at $\omega_\mathrm{c} \equiv \omega_\mathrm{pl}$, the reduced density matrix $\rho^{(\mu)}$ obeys
\begin{align}
\dot{\rho}^{(\mu)}
&=-\mathrm{i}[H^{(\mu)},\rho^{(\mu)}]+\Gamma_{\mu}\,\mathcal D[\hat{\sigma}^{(\mu)}]\rho^{(\mu)},\label{eq:ME_mu_app}\\
H^{(\mu)}
&=\delta_{\mu}\,\ket{\mu}\bra{\mu}+f_{\mu}\left(\hat{\sigma}^{(\mu) \dagger}+\hat{\sigma}^{(\mu)}\right),
\label{eq:H_mu_app}
\end{align}
where $\delta_{\mu}\equiv \omega_{\mu}-\omega_\mathrm{c}$ and, in our symmetric setup, $\delta_{\pm}=\pm \Delta/2$.

Let $p_{\mu}\equiv \rho^{(\mu)}_{\mu\mu}$ and $s_{\mu}\equiv \rho^{(\mu)}_{\mu\mathrm g}$. From \cref{eq:ME_mu_app} one obtains the optical Bloch equations
\begin{align}
\dot p_{\mu}&=-\Gamma_{\mu} p_{\mu}+\mathrm{i} f_{\mu}\left(s_{\mu}^*-s_{\mu}\right),\label{eq:OBE_p_app}\\
\dot s_{\mu}&=-\left(\frac{\Gamma_{\mu}}{2}+\mathrm{i}\delta_{\mu}\right)s_{\mu}
+\mathrm{i} f_{\mu}\left(2p_{\mu}-1\right). \label{eq:OBE_s_app}
\end{align}
Setting $\dot s_{\mu}=0$ gives
\begin{equation}
s_{\mu}=\mathrm{i} f_{\mu}\,\frac{2p_{\mu}-1}{\frac{\Gamma_{\mu}}{2}+\mathrm{i}\delta_{\mu}},
\label{eq:s_in_terms_of_p_app}
\end{equation}
which implies
\begin{equation}
\mathrm{i} f_{\mu}\left(s_{\mu}^*-s_{\mu}\right)=
\Gamma_{\mu}\,\frac{4 f_{\mu}^2}{\Gamma_{\mu}^2+4\delta_{\mu}^2}\left(1-2p_{\mu}\right).
\label{eq:coh_to_pop_app}
\end{equation}
Inserting \cref{eq:coh_to_pop_app} into $\dot p_{\mu}=0$ yields
\begin{equation}
p_{\mu}=\frac{4 f_{\mu}^{2}}{\Gamma_{\mu}^{2}+4\delta_{\mu}^{2}+8 f_{\mu}^{2}},
\label{eq:p_mu_app}
\end{equation}
and substituting back into \cref{eq:s_in_terms_of_p_app} gives the stationary coherence
\begin{equation}
s_{\mu}=-\frac{2 f_{\mu}\,\left(2\delta_{\mu}+\mathrm{i}\Gamma_{\mu}\right)}{\Gamma_{\mu}^{2}+4\delta_{\mu}^{2}+8 f_{\mu}^{2}}.
\label{eq:s_mu_app}
\end{equation}

The factorized structure gives the steady-state matrix elements in the global basis:
\begin{align}
\rho_{++}&=p_{+}(1-p_{-}),\qquad
\rho_{--}=p_{-}(1-p_{+}), \nonumber \\
\rho_{+-}&=s_{+}\,s_{-}^{*},
\label{eq:rho_elements_sec_app}
\end{align}
so that the coherences $\rho_{+-}$ and $\rho_{-+} = \rho_{+-}^*$ are generally nonzero even though
$\rho_{\mathrm{ss}}^{(\mathrm{sec})}$ factorizes.

For the symmetric case, $\Gamma_\mu \equiv \Gamma$ according to \cref{eq:Gamma_pm_def}, with $\delta_{\pm}=\pm\Delta/2$ and $f_{+}=f$, $f_{-}=\eta f$,
it is convenient to define
\begin{equation}
D = \Gamma^{2}+\Delta^{2}+8 f^{2}.
\end{equation}
A short algebra using \cref{eq:PbPd_elements_app,eq:p_mu_app,eq:s_mu_app,eq:rho_elements_sec_app} yields
\begin{align}
\langle \hat P_{\mathrm b}\rangle_{\mathrm ss}^{(\mathrm{sec})}\big|_{\eta=1}
&=\frac{4 f^{2}}{D^{2}}\left(2\Gamma^{2}+4 f^{2}\right) ,\nonumber\\
\langle \hat P_{\mathrm d}\rangle_{\mathrm ss}^{(\mathrm{sec})}\big|_{\eta=1}
&=\frac{4 f^{2}}{D^{2}}\left(2\Delta^{2}+4 f^{2}\right) ,
\label{eq:PbPd_sec_bright_app}
\end{align}
for the bright drive, while for the dark drive the two expressions are exchanged,
\begin{align}
\langle \hat P_{\mathrm b}\rangle_{\mathrm ss}^{(\mathrm{sec})}\big|_{\eta=-1}
&=\langle \hat P_{\mathrm d}\rangle_{\mathrm ss}^{(\mathrm{sec})}\big|_{\eta=1} \nonumber \\
\langle \hat P_{\mathrm d}\rangle_{\mathrm ss}^{(\mathrm{sec})}\big|_{\eta=-1}
&=\langle \hat P_{\mathrm b}\rangle_{\mathrm ss}^{(\mathrm{sec})}\big|_{\eta=1}.
\label{eq:PbPd_sec_dark_app}
\end{align}

\subsection{Non-secular leakage.}

The diagonal form \cref{eq:L_ns_app}, expressed in terms of the bright and dark jump operators
$\hat B_{\mathrm b,d} = (\hat B_{+} \pm \hat B_{-})/\sqrt{2}$, and the closed-form steady-state expressions derived below,
assume an unresolved-transition (common-bath) regime $|\Delta| \lesssim \gamma_{\mathrm D}/2$, so that the suppression factor $\kappa_\mathrm{c}$ in \cref{eq:Wleak_explicit} remains close to unity.
In this limit, the bath does not resolve the two Bohr frequencies $\omega_{\pm} = \omega_{\mathrm c} \pm \Delta/2$,
so $J(\omega_{+}) \simeq J(\omega_{-})$ and the off-diagonal Kossakowski element $g$ remains appreciable;
the dissipator is then well diagonalized by the bright (symmetric) and dark (antisymmetric) superpositions.
For $|\Delta| \gg \gamma_{\mathrm D}/2$ (so that $\kappa_\mathrm{c} \to 0$), the cross correlations are suppressed ($g \to 0$) and the natural jump operators approach $\hat B_{\pm}$ instead, so the bright and dark ladder solution we report below is no longer quantitative.

Moreover, in the analytical solutions presented here, we neglect the coherent $b \leftrightarrow d$ mixing induced by $\hat H_{s} = \frac{\Delta}{2}(\hat N_{+} - \hat N_{-})$ when expressed in the $\{\ket{\mathrm b}, \ket{\mathrm d}\}$ basis, an effect that becomes relevant as $\Delta$ increases. Instead, the complete numerical solutions presented in the main text describe the correct physics. This restriction applies only to the closed-form ladder steady states reported below. The component rate equations (and the numerical GKSL integration) remain valid for arbitrary $\Delta$, and for $\Delta \gg \gamma_\mathbf{D}/2$, the non-secular generator continuously approaches the secular one as $g \to 0$.

Therefore, limiting the validity of the following discussion to the case $\Delta \lesssim \gamma_\mathbf{D}/2$, we use \cref{eq:bright_dark,eq:B_bright_dark_BR} to write the Hamiltonian for the bright and dark drives as
\begin{align}
\hat H_{\mathrm{drive}}^{(\eta=1)}&=\sqrt2 f\left(\hat B_{\mathrm b}+\hat B_{\mathrm b}^{\dagger}\right),\qquad \text{bright drive} \nonumber \\
\hat H_{\mathrm{drive}}^{(\eta=-1)}&=\sqrt2 f\left(\hat B_{\mathrm d}+\hat B_{\mathrm d}^{\dagger}\right), \qquad \text{dark drive}.
\label{eq:Hdrive_bd_ns_app}
\end{align}
For a given drive choice $\eta$, the master equation is
\begin{equation}
\dot{\hat\rho}=-\mathrm{i}[\hat H_{\mathrm{drive}}^{(\eta)},\hat\rho]+\mathcal L_{\mathrm{leak}}^{(\mathrm{ns})}\hat\rho.
\label{eq:ME_ns_app}
\end{equation}
In this subsection it is more useful to work in the basis $\{\ket{\mathrm g},\ket{\mathrm b},\ket{\mathrm d},\ket{\mathrm e}\}$ and use $\rho_{mn}\equiv\bra{m}\hat\rho\ket{n}$ with $m,n\in\{\mathrm g,\mathrm b,\mathrm d,\mathrm e\}$.

\paragraph*{Bright drive.}
For $\eta=1$, the Hamiltonian couples the resonant ladder
$\ket{\mathrm g}\leftrightarrow\ket{\mathrm b}\leftrightarrow\ket{\mathrm e}$, while $\ket{\mathrm d}$ is populated only via the decay $\ket{\mathrm e}\to\ket{\mathrm d}$ generated by $\hat B_{\mathrm d}$.
The equations of motion for the populations and the coherences needed to close the system are
\begin{align}
\dot\rho_{\mathrm{dd}}&=\Gamma_{\mathrm d}\,(\rho_{\mathrm{ee}}-\rho_{\mathrm{dd}}),\label{eq:dddot_ns_app}\\
\dot\rho_{\mathrm{ee}}&=-(\Gamma_{\mathrm b}+\Gamma_{\mathrm d})\,\rho_{\mathrm{ee}}
-\mathrm{i}\sqrt2  f\,(\rho_{\mathrm{be}}-\rho_{\mathrm{eb}}),\label{eq:eedot_ns_app}\\
\dot\rho_{\mathrm{bb}}&=-\Gamma_{\mathrm b}\,(\rho_{\mathrm{bb}}-\rho_{\mathrm{ee}})
-\mathrm{i}\sqrt2 f\,(\rho_{\mathrm{gb}}-\rho_{\mathrm{bg}}-\rho_{\mathrm{be}}+\rho_{\mathrm{eb}}),\label{eq:bbdot_ns_app}\\
\dot\rho_{\mathrm{gb}}&=-\frac{\Gamma_{\mathrm b}}{2}\rho_{\mathrm{gb}}+\Gamma_{\mathrm b}\rho_{\mathrm{be}}
+\mathrm{i}\sqrt2 f\,(\rho_{\mathrm{gg}}+\rho_{\mathrm{ge}}-\rho_{\mathrm{bb}}),\label{eq:gbdot_ns_app}\\
\dot\rho_{\mathrm{be}}&=-\Bigl(\Gamma_{\mathrm b}+\frac{\Gamma_{\mathrm d}}{2}\Bigr)\rho_{\mathrm{be}}
-\mathrm{i}\sqrt2 f\,(\rho_{\mathrm{ge}}-\rho_{\mathrm{bb}}+\rho_{\mathrm{ee}}),\label{eq:bedot_ns_app}\\
\dot\rho_{\mathrm{ge}}&=-\frac{\Gamma_{\mathrm b}+\Gamma_{\mathrm d}}{2}\rho_{\mathrm{ge}}
+\mathrm{i}\sqrt2 f\,(\rho_{\mathrm{gb}}-\rho_{\mathrm{be}}),\label{eq:gedot_ns_app}
\end{align}
with $\rho_{\mathrm{gg}}=1-\rho_{\mathrm{bb}}-\rho_{\mathrm{dd}}-\rho_{\mathrm{ee}}$.
At steady state, \cref{eq:dddot_ns_app} gives $\rho_{\mathrm{dd}}=\rho_{\mathrm{ee}}$; eliminating the coherences from \cref{eq:eedot_ns_app}--\cref{eq:gedot_ns_app}
yields two algebraic equations for $\rho_{\mathrm{bb}}$ and $\rho_{\mathrm{ee}}$, whose solution can be written compactly as
\begin{equation}
\label{eq:ss_b_drive_ns_app}
\rho_{\mathrm{bb}}^{(\mathrm{ns})} =\frac{32 f^{2}\Bigl[\Gamma^2 + 2 f^{2}\Bigr]}{\mathcal N_{\mathrm b}} , \qquad \rho_{\mathrm{dd, ee}}^{(\mathrm{ns})} =\frac{64 f^{4}}{\mathcal N_{\mathrm b}}  
\end{equation}
where $\mathcal N_{\mathrm b} = 4 \Gamma^2 \bigl(\Gamma_{\mathrm b}^{2}+16 f^{2}\bigr) + (4f)^4$, having also exploited $\Gamma_{\mathrm b}+\Gamma_{\mathrm d} = 2 \Gamma$. 
Since $\hat P_{\mathrm b}=\ket{\mathrm b}\bra{\mathrm b}$ and $\hat P_{\mathrm d}=\ket{\mathrm d}\bra{\mathrm d}$ in the single-excitation manifold, one has
$\langle \hat P_{\mathrm b}\rangle_{\mathrm ss}^{(\mathrm{ns})}=\rho_{\mathrm{bb}}^{(\mathrm{ns})}$ and
$\langle \hat P_{\mathrm d}\rangle_{\mathrm ss}^{(\mathrm{ns})}=\rho_{\mathrm{dd}}^{(\mathrm{ns})}$.

\paragraph*{Dark drive.}
For $\eta=-1$, the same derivation applies upon exchanging $\mathrm b\leftrightarrow \mathrm d$, i.e.
\begin{equation}
\label{eq:ss_d_drive_ns_app}
\rho_{\mathrm{bb, ee}}^{(\mathrm{ns})} =\frac{64 f^{4}}{\mathcal N_{\mathrm b}} , \qquad \rho_{\mathrm{dd}}^{(\mathrm{ns})} =\frac{32 f^{2}\Bigl[\Gamma^2 + 2 f^{2}\Bigr]}{\mathcal N_{\mathrm d}} ,
\end{equation}
where $\mathcal N_{\mathrm d} = 4 \Gamma^2 \bigl(\Gamma_{\mathrm d}^{2}+16 f^{2}\bigr) + (4f)^4$, so that $\langle \hat P_{\mathrm b}\rangle_{\mathrm ss}^{(\mathrm{ns})}=\rho_{\mathrm{bb}}^{(\mathrm{ns})}$ and
$\langle \hat P_{\mathrm d}\rangle_{\mathrm ss}^{(\mathrm{ns})}=\rho_{\mathrm{dd}}^{(\mathrm{ns})}$.

\clearpage


%


\end{document}